\documentclass[preprint,10pt,1p]{elsarticle}
\newcommand{\sgn}{\text{sgn}}
\biboptions{sort&compress}
\usepackage{color}
\usepackage{amsmath}
\usepackage{amssymb}

\usepackage{graphicx}
\usepackage[figuresright]{rotating}

\def\PD#1#2{\frac{\partial #1}{\partial #2}}

\def\BS#1{{\bf #1}}

\journal{Journal of Computational Physics}
\begin{document}
\begin{frontmatter}
\title{Simulating hard rigid bodies.}
\author{C. De~Michele \corref{CDM}}
\address{Dipartimento di Fisica, Universit\`a di Roma ``La Sapienza'',\\ P.le Aldo Moro 2, 00185 Roma, Italy}%
\ead{cristiano.demichele@roma1.infn.it}
\cortext[CDM]{Tel.: +390649913524; fax: +39064463158.} 



\begin{abstract}
Several physical systems in condensed matter have been modeled approximating
their constituent particles as hard objects. The hard spheres 
model has been indeed one of the cornerstones of the computational and theoretical 
description in condensed matter. The next level of description is to consider 
particles as rigid objects of generic shape, which would enrich the possible 
phenomenology enormously. This kind of modeling will prove to be interesting 
in all those situations in which steric effects play a relevant role. These include biology, 
soft matter, granular materials and molecular systems.
With a view to developing a general recipe for event-driven Molecular Dynamics 
simulations of hard rigid bodies, two algorithms for calculating the distance between two convex 
hard rigid bodies and the contact time of two colliding hard rigid bodies 
solving a non-linear set of equations will be described. 
Building on these two methods, an event-driven 
molecular dynamics algorithm for simulating systems of convex hard rigid bodies
will be developed and illustrated in details.  In order to optimize the 
collision detection between very elongated hard rigid bodies, a novel 
nearest-neighbor list method based on an oriented bounding box will be introduced 
and fully explained. Efficiency and performance of the new algorithm proposed 
will be extensively tested for uniaxial hard ellipsoids and superquadrics.
Finally applications in various scientific fields will be reported and discussed.

\vspace{1pc}
\end{abstract}
\end{frontmatter}
\section{Introduction}


Systems which are composed of many particles can often be modeled as 
an ensemble of hard rigid bodies. Such description is particularly successful
when excluded volume interactions are dominant and  internal vibrational
degrees of freedom are negligible. Despite the absence
of any attraction, particles interacting with only excluded volume interactions
exhibit a rich phase diagram with a multiplicity of phases, especially when the shape is
a non-spherical one\cite{AllenReview,FrenkelMulderMolPhys85,Allen93,TalbotKivelsonAllen,AllenWarren}.

Hard Spheres (HS) are a classical example of hard-body model, which has been particularly
useful to understand the basic correlation which develops in simple fluids \cite{GrayGubbinsBook,WCAACP76,WCAjcp71, HansenMcDonaldBook,kolnezMolPhys87} and provides hints on the
slow dynamics which characterize liquids approaching the glass transition, where packing effects become
even more significant \cite{PuseyChapter,foffiPRL2003,foffiPRE2004}.
Even in the case of molecular fluids, HS models are a good starting point for  sophisticated liquid matter theories  \cite{GrayGubbinsBook}. 

Non-spherical models of rigid bodies are crucial to understand 
the role of the rotational degrees of freedom, as well as the role  played by the shape 
in determining the system's physical properties.
Onsager, introducing a hard sphero-cylinder model (HSC) for liquid crystals,
showed that, by only changing the aspect ratio of the particles, a nematic phase
can become thermodynamically stable \cite{Onsager49}. In Onsager's theory, the internal energy of the system is zero and 
only the entropy, coming from translational and rotational degrees of freedom of the particles, contributes to the free energy of the system.
Systems showing an ``entropically driven'' phase transition have been  
extensively studied over the last $60$ years \cite{DeGennes95,Vroege92}.
The  study of  such transitions has been  based  on extensions of the 
 original Onsager's theory \cite{Parsons,Lee2,allenHEPhaseDiag,MargoEvans}, 
 and complemented by  experiments  (for a recent review see \cite{Vroege92}).
 
Most of the information for hard-body systems has been calculated using numerical 
simulations \cite{allenHEPhaseDiag,Frenkel84}. Several numerical techniques have been developed in the
past to simulate particles interacting with only excluded volume interactions. The essence of these numerical
algorithms involves the evaluation of the overlap between different objects or, equivalently, their geometrical
distance. The first  simulations of HS \cite{Alder57} were carried out by 
  Alder and Wainwright in 1957, and they provided the first evidence of a crystal phase in the case of
  spherical hard particles (disks and spheres). In 1972 Vieillard-Baron \cite{Vieillard72}
published  a numerical investigation of the phase diagram of a two-dimensional hard-ellipsoids (HEs) fluid, 
introducing an overlap criterion for HEs suitable for a Monte-Carlo (MC) simulation.
Building on the work of Viellard-Baron, Frenkel {\it et al.} \cite{Frenkel84} investigated 
in 1984 the phase diagram of a tridimensional system of HEs, through MC simulations. 
 Perram and Wertheim \cite{Perram85} introduced a simpler overlap criterion, which has been
recently used by Donev {\it et al.} to perform molecular dynamics simulations
of HE \cite{Donev05a,Donev05b,DonevScience} and which has been also generalized 
to any couple of smooth convex shapes \cite{DonevThesis06,DonevPRB06,JiaoPRL08,JiaoPRE09}.
Simulations of particles with more complex shapes have also been reported. 
For example in 1986 Stroobants {\it et al.} carried out MC simulations of a system of hard sphero-cylinders (HSC), i.e. molecules consisting of a hard cylindrical rod of length L and diameter D, capped at each end by hard hemispheres also of diameter D. These simulations are computationally less demanding \cite{FrenkelBook} than the HE ones.
More recently, MC simulations of hard-cylinders (HCY) have been performed \cite{Blaak99} in order to look for a cubatic phase,  which had been reported for cut hard spheres \cite{Veerman92}.   
The oblate version of HSCs (i.e. the solid resulting from the intersection of two spheres) has been investigated by MC simulations under the name of UFO (the name comes from the shape of the particles) \cite{Het99}.

Molecular dynamics simulations of hard bodies are less common than the Monte Carlo ones, 
since the implementation of the 
overlap criterion between hard bodies must be complemented with an algorithm estimating the
collision time between them. The evolution of the system in configuration space is propagated
from one collision to the next, giving rise to the so-called event-driven molecular dynamics (EDMD). 
In EDMD the predicted collision times between hard bodies are computed and stored into a time-ordered event calendar (as it was first done for HSs by Rapaport \cite{RapaBook},
although different techniques also exist in literature \cite{lubachevskyED}). Such technique has been recently extended to Brownian Dynamics of HSs \cite{BD4HS,donevBD}.

A basic scheme for simulating hard non-spherical bodies is based on the standard MD algorithms, where at the end of each time step a check for possible overlaps is performed and the simulation is ``rewinded" when an overlap occurs \cite{Allen89,AllenPRLHE}. This scheme is obviously inefficient.  
An overlap potential of two hard bodies $A$ and $B$ is a function $F(A,B)$, such that $F<0$  if $A$ and $B$ overlap. 
In general it is not easy (and not necessarily possible) to find an analytic form for the overlap potential of two rigid bodies of arbitrary convex shape, except for the aforementioned cases of UFO, HSC, HCY, HCS and HE. 
Overlap potentials for hard rigid convex bodies can be found in \cite{Allen93}.

An EDMD algorithm for non-spherical objects employing an event-calendar has been recently proposed by Donev {\it et al.} \cite{Donev05a} and tested on several systems of hard particles \cite{Donev05b,DonevPRB06,JiaoPRL08,JiaoPRE09} .  Such algorithm \cite{Donev05a,Donev05b} requires the use of overlap potentials \cite{Perram85,Vieillard72}, like in MC simulations. 
In the present paper a different route to provide a general algorithm to simulate hard particles will be presented.

If the surface of two hard rigid bodies (HRB) is smooth enough (i.e. first and second
order derivatives are defined over the whole surface) a possible overlap
potential is provided by the minimum distance between the two surfaces (defining  the distance as negative
when two rigid bodies overlap). In this paper we illustrate a method to calculate the distance between two generic convex HRBs based on a Newton-Raphson (NR) method. 
We will also illustrate a simple algorithm to efficiently evaluate 
a guess 
 of the collision contact point and time. Starting from this guess true collision point and time
 can be calculated by solving a reduced system of equations, again through 
a Newton-Raphson method. 
It is worth noting that this algorithm can handle grazing collisions, i.e. collisions in which
two rigid bodies touch slightly \cite{Donev05b}, without  
tuning the algorithm parameters significantly. 
Furthermore,  in the case of HEs and superquadrics (SQs),  we illustrate a new kind of neighbour list based on oriented bounding boxes that can also be easily generalized to more complex shapes. 
Like the algorithm proposed by Donev {\it et al.}  \cite{JiaoPRL08,JiaoPRE09}, 
our method can be easily generalized to arbitrary convex shapes 
(but also decorated with localized patchy interactions, see below) with similar efficiency.
 
The present algorithm has been already applied successfully to the simulation and to the study of various systems.  
It has been implemented in the study of structural and dynamical properties
of uniaxial HE \cite{ourHEstatic,DeMicheleNEMGLASS}. With this code,  adding the possibility of having
localized patchy square-well interactions,  it has been possible to study the statics and the dynamics of primitive models of Water \cite{DeMicheleSticky06} and Silica \cite{DeMicheleSilica06}, as well as the irreversible gelation of a model inspired by stepwise polymerization of bifunctional diglycidyl-ether of bisphenol-A with pentafunctional diethylenetriamine \cite{DeMicheleDGEBA08,corezziJPCB,DeMicheleDouglas08}. 
More recently this code has also been generalized in order to simulate a coarse-grained model of biological systems and primitive models of proteins. 

In Section \ref{sec:EMD4RB} we introduce the general algorithm for the EDMD of rigid bodies.  This requires
discussing the HRB motion, the evaluation of the 
distance and the time of collision among HRBs  and the optimized  linked list method.
In Section \ref{sec:EMD4HE} we specialize the results of Section \ref{sec:EMD4RB} to the specific case of HEs; in particular, we discuss a new nearest-neighbours list method, that over-performs the simple linked lists method usually employed in EDMD in the case
of very elongated HRBs.
In Section \ref{sec:perform} the performance in the case of HEs at various densities, elongations and with respect to the main parameters of EDMD is analyzed, while in Section \ref{sec:supquad} the performance of the algorithm in the case of SQs 
is investigated.
In Section \ref{Sec:application} some perspectives and applications of this new algorithm are discussed,
and in Section \ref{Sec:conclusions} conclusions are drawn.
\section{An event-driven algorithm for rigid bodies.}\label{sec:EMD4RB}

\subsection{Geometry of rigid bodies.}

The orientation of a HRB can be represented by the $3$ column eigenvectors ${\bf u}_i$ (with $i=1,2,3$) of the  inertia tensor expressed in the laboratory reference system. These vectors form an orthogonal set and can be arranged in a matrix ${\bf R}$, i.e. 
\begin{equation}
{\bf R} =  ( {\bf u}_0\  {\bf u}_1\  {\bf u}_2 )^T
\end{equation}
where $A^T$ indicates the transpose of $A$.

This matrix will be referred to as the ``orientational matrix'' in the following. The orientational matrix relates the coordinates ${\bf x}$ in the laboratory reference system to  the coordinates ${\bf x'}$ in the HRB reference system via:
\begin{equation}
{\bf x}' = {\bf R} {\bf x}
\end{equation}

The following discussion focuses on HRBs with finite volume and bounded surface. We assume that the equation of the surface of the HRB, in the ``HRB reference system'' with origin in the center of mass and axes parallel to the vectors $\{{\bf u}_i\}_i$, is of the form $f(\bf{r})=0$, where $f$ changes sign passing from the interior to the exterior of the HRB.
Moreover, the normal $\frac{\partial f}{\partial {\bf r} }$  and its first order derivatives are assumed to be properly defined.

\subsection{Motion of rigid bodies}\label{ssec:rotation}

In the following equations we assume that the three eigenvalues of the inertia tensor are all equal to $I$.
The formulas for the free rotation of a symmetric-top case are just slightly more elaborated \cite{LandauMecBook}, although the free rotation for a general rigid body involves the calculation of special functions \cite{LandauMecBook,WhittakerBook} and requires a more sophisticated algorithm which has been implemented only recently \cite{VanZonJCP08,DeLaPena07}. 
Nevertheless, this paper being focused on an algorithm which predicts collision between 
HRBs, for the sake of simplicity the discussion will be restricted to the case of a fully symmetric inertia tensor.
It is straightforward to generalize the approach to a completely asymmetric inertia tensor.

From  the angular velocity ${\bf w}=(w_x,w_y,w_z)$ of a free rigid body the antisymmetric matrix $\bf \Omega$ can be built, i.e.:
\begin{equation}
{\bf \Omega} = \begin{pmatrix}0 & -w_z & w_y\cr w_z & 0 & -w_x\cr -w_y & w_x & 0\end{pmatrix}
\label{Eq:omegmat}
\end{equation}

It is possible to relate the orientation ${\bf R}(t)$ to the orientation at time $t=0$  via \cite{LandauMecBook,GoldsteinBook}:
\begin{equation}
{\bf R}(t) = {\bf R}(0) ({\bf I} + {\bf M})
\label{Eq:Rt}
\end{equation}
where ${\bf M}$ is the following matrix:
\begin{equation}
{\bf M} = - \frac{\sin(wt)}{w} {\bf \Omega} + \frac{1-\cos(wt)}{w^2} {\bf \Omega}^2
\label{Eq:Mmat}
\end{equation}
with $w = \|{\bf w}\|$. Note that if $w=0$ then ${\bf R}(t) = {\bf R}(0)$.


The update of position and orientation of a free rigid body is therefore accomplished by:
\begin{subequations}
\label{Eq:surf}
\begin{equation}
{\bf x}(t) = {\bf x}(0) + {\bf v} t 
\label{Eq:surfa}
\end{equation}
\begin{equation}
{\bf R}(t) = {\bf R}(0) ({\bf I} + {\bf M})
\label{Eq:surfb}
\end{equation}
\end{subequations}
where ${\bf x}(t)$ is the position of the center of mass of the rigid body at time $t$ and $\bf v$ is its velocity.

\subsection{Distance between two rigid bodies}
\begin{figure}
\begin{center}
\includegraphics[width=0.49\linewidth]{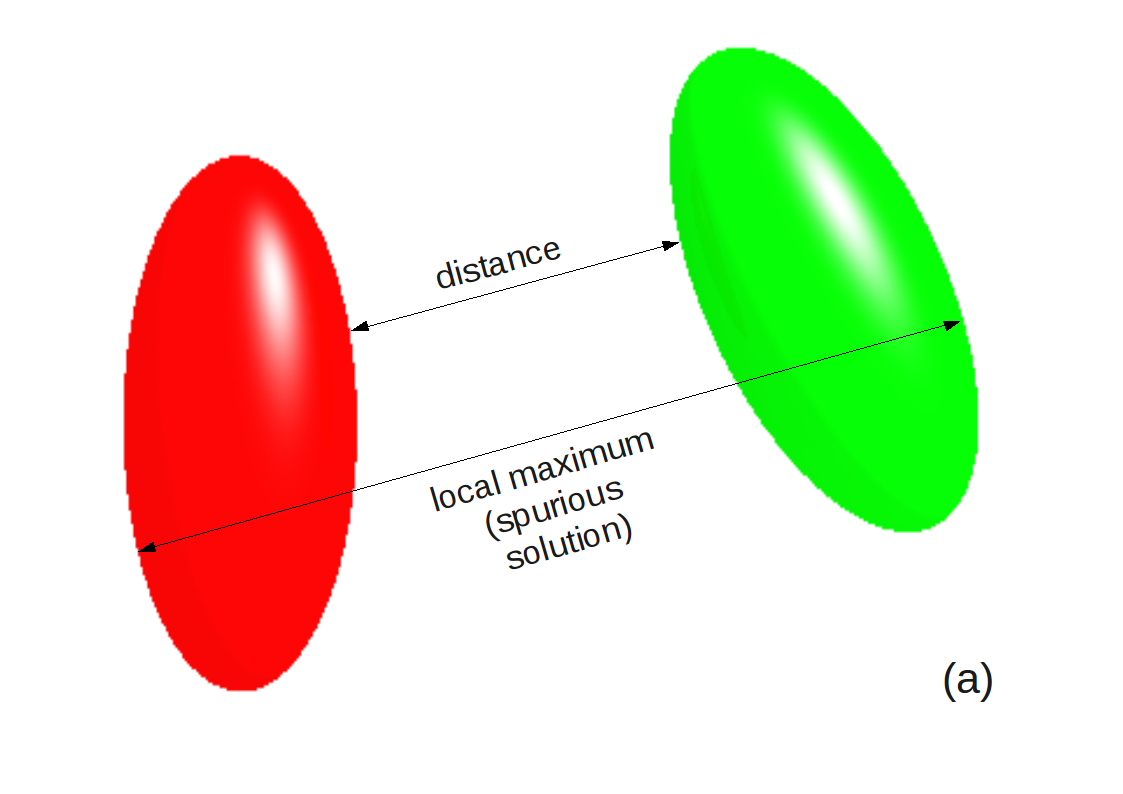}
\includegraphics[width=0.49\linewidth]{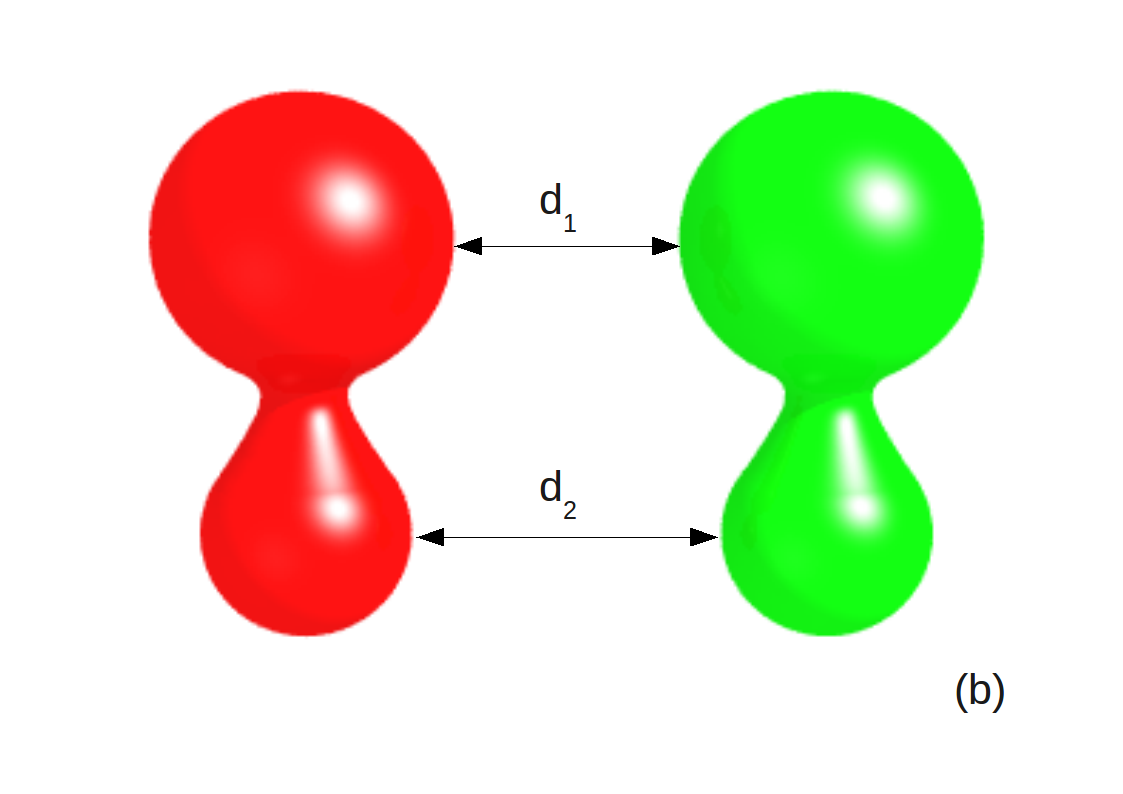}
\caption{\label{fig:many_d} Examples of solution of Eq. \ref{Eq:dist8}. (a) Distance between tow HRBs with
a possible spurious solution. 
(b) For non-convex objects, there can be multiple solutions.}
\end{center}
\end{figure}

The present algorithm is based on a calculation of the distance between HRBs. 
In the following we will show how such a distance is calculated.
Consider two rigid bodies, $A$ and $B$, whose surfaces are implicitly defined by the following equations:
\begin{subequations} \label{Eq:shapes}
\begin{equation}\label{Eq:shapef}
f({\bf x}) = 0
\end{equation}
\begin{equation}\label{Eq:shapeg}
g({\bf x}) = 0
\end{equation}
\end{subequations}
It is also assumed that if a point ${\bf x}$ is inside the rigid body $A$ then $f({\bf x}) < 0$  ($g({\bf x}) < 0$),  while if it is outside $f({\bf x}) > 0$ ($g({\bf x}) > 0$). The distance $d$ between these two objects can be defined as follows:
\begin{equation}\label{Eq:distdef}
d = \min_{{f({\bf x}_A)=0}\atop{g({\bf x}_B)=0}} \| {\bf x}_A - {\bf x}_B \|
\end{equation}
The latter equation means that the quantity $ \| {\bf x}_A - {\bf x}_B \|$ has to be 
minimized with the constraints that points ${\bf x}_A$ and ${\bf x}_B$ 
belong respectively to the surfaces of $A$ and $B$.
Hence, introducing the Lagrange multipliers $\alpha$ and $\beta$,
the distance between the two HRBs $A$ and $B$ can be defined as the solution 
of the following set of $8$ equations:
\begin{subequations} \label{Eq:dist8}
\begin{equation} \label{Eq:dist8a}
f_{{\bf x}_A} = - \alpha^2 g_{{\bf x}_B}
\end{equation}
\begin{equation} \label{Eq:dist8b}
f({\bf x}_A) = 0
\end{equation}
\begin{equation} \label{Eq:dist8c}
g({\bf x}_B) = 0
\end{equation}
\begin{equation} \label{Eq:dist8d}
 {\bf x}_A + \beta f_{{\bf x}_A}={\bf x}_B
\end{equation}
\end{subequations}
where ${\bf x}_A = (x_A,y_A,z_A)$, ${\bf x}_B = (x_B,y_B,z_B)$,
$f_{{\bf x}_A} = \left (\PD{f}{{\bf x}_A}\right )^T$ and $g_{{\bf x}_B} = \left (\PD{g}{{\bf x}_B}\right )^T$\footnote{Note that 
the gradient $\PD{}{\bf x}$ is intended to be a row vector.}.

Eqs. (\ref{Eq:dist8b}) and (\ref{Eq:dist8c}) guarantee that ${\bf x}_A$ and ${\bf x}_B$ are points on $A$ and $B$, Eq.(\ref{Eq:dist8a}) guarantees that the normals to the surfaces are anti-parallel, and Eq.(\ref{Eq:dist8d}) guarantees that the displacement of ${\bf x}_A$ from ${\bf x}_B$ is collinear to the normals to the surfaces.

Equations (\ref{Eq:dist8}) define extremal points of $d$ \cite{TaylorMannBook}; for example for two convex 
HRBs local maxima can be found (see Fig. \ref{fig:many_d}(a)) and 
for two general non-overlapping non-convex HRBs these equations can have multiple solutions (see Fig. \ref{fig:many_d}(b)), although only the smallest one is the actual distance. Therefore, to solve these equations iteratively it is necessary to start from a good initial guess of $\left( {\bf x}_A,{\bf x}_B,\alpha,\beta \right )$ to avoid finding spurious solutions.

\begin{figure}
\begin{center}
\includegraphics[width=0.65\linewidth]{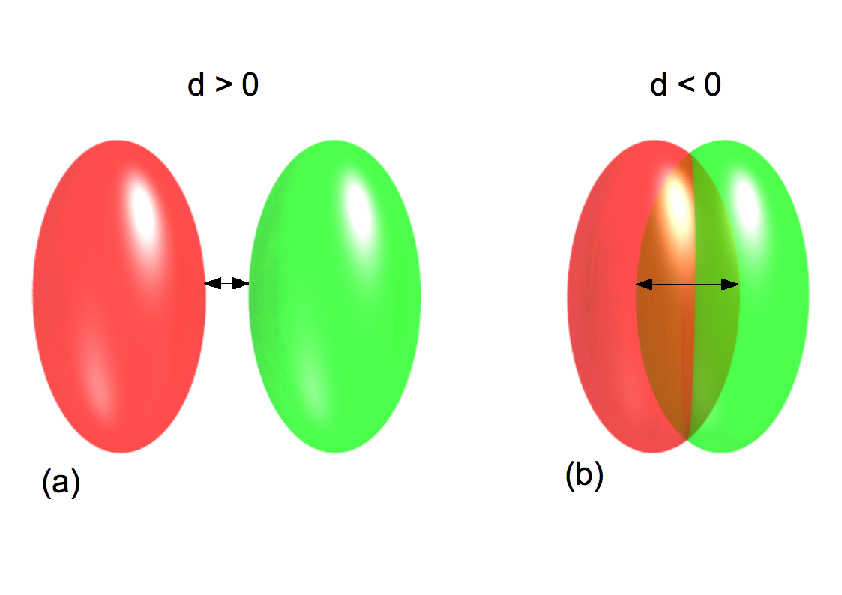}
\caption{\label{fig:negative_d} When convex objects slightly overlap, the solution of Eqs.(\ref{Eq:dist8}) changes sign.}
\end{center}
\end{figure}

In addition we note that if two rigid bodies overlap slightly (i.e. the overlapped volume is small) there is a solution with $\beta < 0$, that is a measure of the inter-penetration of the two rigid bodies;  such a solution will be referred to as the ``negative distance'' solution (Fig.\ref{fig:negative_d}). 

Finally, we define the quantities $d_i$ as follows:
\begin{equation}
d_i \equiv \| {\bf x}_A^i - {\bf x}_B^i \|
\end{equation}
where $({\bf x}_{A}^i,{\bf x}_B^i, \alpha_i, \beta_i)$ is a solution of Eqs. (\ref{Eq:dist8}), 
the distance $d$ between two HRBs can be formally written as:
\begin{equation}
d = \sgn(\beta_{min})\ \min_i\{d_i\}
\label{eq:distdef}
\end{equation}
where $\sgn(x)$ is the sign function and $\beta_{min}$ is the $\beta_i$ corresponding to the 
solution with the smallest $d_i$.

\subsubsection{The Newton-Raphson method for calculating the distance}

The set of equations (\ref{Eq:dist8}) can be conveniently solved by a Newton-Raphson (NR) method
\cite{NumRecipes}. This method, as long as a good initial guess is provided,  reaches  the solution very quickly thanks to its quadratic convergence \cite{NumRecipes}. If one defines:
\begin{equation}
{\bf F}({\bf y})=
\begin{pmatrix}
f_{{\bf x}_A} + \alpha^2 g_{{\bf x}_B}\cr
f({\bf x}_A)\cr
g({\bf x}_B)\cr
{\bf x}_A + \beta f_{{\bf x}_A}-{\bf x}_B
\end{pmatrix}
\end{equation}
The Eqs. (\ref{Eq:dist8}) become:
\begin{equation}\
{\bf F}({\bf y})= 0
\end{equation}
where ${\bf y}=({\bf x}_A, {\bf x}_B, \alpha, \beta)$.

Given an initial point ${\bf y}_0$, a sequence of points converging to the solution can be built as follows:
\begin{equation}
{\bf y}_{i+1} = {\bf y}_i + {\bf J}^{-1} {\bf F}({\bf y}_i)
\label{Eq:NRiter}
\end{equation}
where ${\bf J}$ is the Jacobian of ${\bf F}$, i.e.:
\begin{equation}
{\bf J} = 
\begin{pmatrix} \PD{f_{\bf x_A}}{\bf x_A}&
\alpha^2\PD{g_{\bf x_B}}{\bf x_B}& 
2 \alpha g_{\bf x_B} & {\bf 0} \cr
{f}_{\bf x_A}^T & {\bf 0}^T & 0 & 0\cr
{\bf 0}^T & {g}_{\bf x_B}^T & 0 & 0\cr
{\bf I} + \beta\PD{f_{\bf x_A}}{\bf x_A} & -{\bf I} & 0 & f_{\bf x_A}\cr
\end{pmatrix}
\label{Eq:JacobDist8}
\end{equation}

\noindent with ${\bf I}$ being the identity $3\hbox{x}3$ matrix and $\bf 0$ the null column vector.

The NR method may not converge if the initial guess is not close enough to the root, hence in general  it may be convenient 
for the sake of numerical robustness to use a globally convergent NR method (see \cite{NumRecipes})
that converges to the solution from any starting point. An alternative route to ensure the 
appropriate robustness is to provide an accurate initial guess, e.g. making use of a steepest descent method, 
as it will be shown in the next section.  
The matrix inversion, required to evaluate ${\bf J}^{-1}$, can be performed by standard $LU$  decomposition \cite{NumRecipes}.  $LU$ decomposition is of order $N^3/3$, where $N$ is the number of equations ($8$ in the present case).  In Sec. \ref{ssec:reducesystem} it will be shown that the above set of equations can be reduced to $5$, thus reducing the time to invert the matrix  by a factor $\sim 4$.

\subsubsection{Initial Guess for the distance: the steepest-descent method.}
\label{sec:iniguess}

\begin{figure}
\begin{center}
\includegraphics[width=0.7\linewidth]{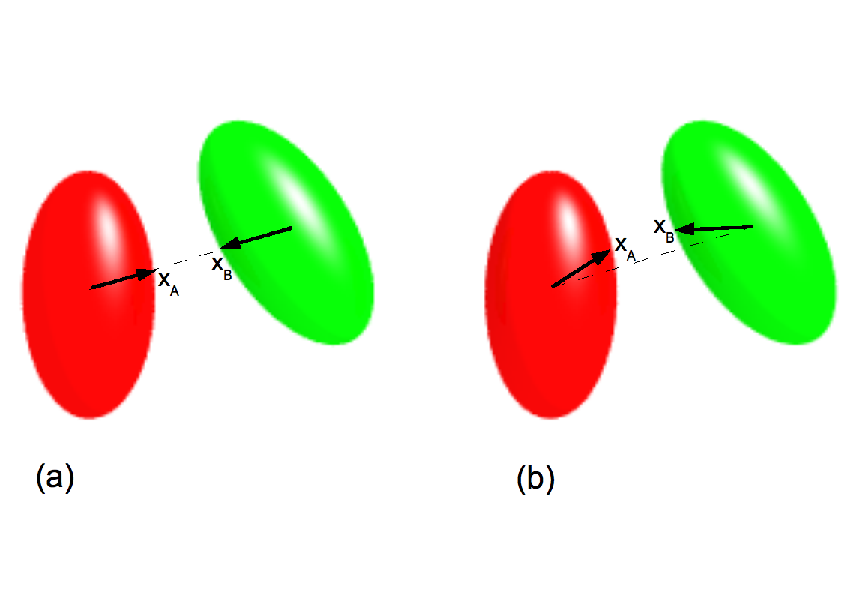}
\vspace{-1.5cm}
\caption{\label{fig:ABguess} Initial guess for the distance to be used as a starting point for NR method.
(a) Simple initial guess for the distance: initial points ${\bf x}_A$
and ${\bf x}_B$ for the NR are obtained considering the interceptions of the line
joining the two HEs centers with their surfaces. (b) A better guess for the distance in the case of moderately elongated HRBs (see discussion in Sec. \ref{ssec:HomoGuess} for the HEs case).}
\end{center}
\end{figure}

As initial guess of the NR it may be convenient to choose the closest pair of points on the intersection of the surfaces with the line joining the center of mass of the two HRBs (Fig.~\ref{fig:ABguess}a). Such guess is reasonable if the objects are almost spherical; otherwhise a refinement is needed.

A general method to refine such initial guess is to minimize the function
\begin{equation}
D({\bf y}) = \alpha_{SD} \| {\bf x}_A - {\bf x}_B \|^2
\label{Eq:distSD}		
\end{equation}
where $\BS y=({\bf x}_A,{\bf x}_B)$ with the constraints $f({\bf x}_A) = 0$ and $g({\bf x}_B) = 0$,
and $ \alpha_{SD}$ is parameter that can be tuned to optimize the steepest-descent (SD).
For solving such a problem, steepest descent steps are used, followed by corrections that hold the points ${\bf x}_A$ and ${ \bf x}_B$  on the surface of the two bodies. The algorithm is the following one:
\begin{enumerate}
\item Choose an initial guess for ${ \bf x}_A$ and $ {\bf x}_B$
\item Evaluate the gradient $D_{\bf y}$ of $D({\bf y})$:
\begin{equation}
 D_{\bf y} = (\BS h_A, \BS h_B ) = 2\alpha_{SD} ({\bf x}_A - {\bf x}_B,  -({\bf x}_A - {\bf x}_B))
\end{equation} 
\item  Calculate the components of $\BS h_A$ and $\BS h_B$ parallel to the surface, i.e.:
\begin{equation}
D_{\bf y}^{\parallel} = (\BS h_A^{\parallel},\BS h_B^{\parallel})
\end{equation}
where 
\begin{equation}
\BS h_\mu^{\parallel} =  \BS h_\mu - (\BS h_\mu\cdot{\bf\hat n_\mu}) {\bf\hat n_\mu}
\end{equation}
with $\mu \in \{A,B\}$ and $\hat{\BS n}_A = f_{{\bf x}_A}/\|f_{{\bf x}_A}\|$, $\hat{\BS n}_B = g_{{\bf x}_B}/\|g_{{\bf x}_B}\|$
\item Move the two points in the direction of $D_{\bf y}^{\parallel}$:
\begin{equation}
\label{Eq:SDmove}
{\bf y}' = (\BS x_A', \BS x_B') = {\bf y} - \lambda_{SD} D_{\bf y}^{\parallel} 
\end{equation}
whith $0 < \lambda_{SD}< 1$.
\item Add a small displacement $d {\bf y}=(\xi_A f_{\BS x_A}, \xi_B g_{\BS x_B})$ to the vector  ${\bf y}'$, i.e.:
\begin{equation}
\BS y'' = (\BS x_A'', \BS x_B'') =  \BS y' + d {\BS y}
\end{equation}
where $\xi_A$ and $\xi_B$ are such that the constraints are satisfied and the two points ${\bf x}_A''$ and ${\bf x}_B''$ belong to  the surfaces of the two rigid bodies, i.e. 
\begin{subequations}
\begin{equation}
f({\bf x}_A' + \xi_A  f_{\BS x_A}) = 0
\end{equation}
\begin{equation}
g({\bf x}_B' + \xi_B  g_{\BS x_B}) = 0
\end{equation}
\label{Eq:adjxSD}
\end{subequations}
\item Terminate if the angle between ${\bf\hat n}$ and $D_{\bf y}$ is small enough, otherwise  go back to step 1.
\end{enumerate}
This procedure provides a guess for $\BS x_A$ and $\BS x_B$. 
Note that the adjustment of the position of the points $A$ and $B$ to hold them respectively onto the surfaces $f$ and $g$
can be implemented again with a one-dimensional Newton-Raphson method, that will generally converge in few steps if the points are not too far from the surfaces.
 
The NR method for the distance requires also a guess for $\alpha$ and $\beta$ 
(introduced in Eq. (\ref{Eq:dist8}));  
$\alpha = \|f_{\BS x_A}\|/\|g_{\BS x_B}\|$, $\beta = 0$ proved to be a good guess.
If the accuracy required for the convergence of the SD is high enough,
the NR method will always converge to the correct solution.  
However, being the SD method much slower than NR, a trade-off between accuracy and speed is needed. 
\subsubsection{Reduced system of equations}
\label{ssec:reducesystem}
The system in Eq. (\ref{Eq:dist8}) can be reduced to $5$ equations eliminating Eq. (\ref{Eq:dist8d}): 
\begin{subequations}
\label{Eq:dist5}
\begin{equation} \label{Eq:dist5a}
f_{{\bf x}_A} + \alpha^2 g_{{\bf x}_B}({\bf x}_A + \beta f_{{\bf x}_A}) = 0
\end{equation}
\begin{equation} \label{Eq:dist5b}
f({\bf x}_A) = 0
\end{equation}
\begin{equation}\label{Eq:dist5c} 
g({\bf x}_A + \beta f_{{\bf x}_A}) = 0
\end{equation}
\end{subequations}

In this case the Jacobian is:
\begin{equation}
\label{Eq:JacobDist5}
{\bf J} = 
\begin{pmatrix} \PD{f_{\bf x_A}}{\bf x_A}+\alpha^2 {\bf A} &
2\alpha g_{\bf x_B} & \alpha^2 \PD{g_{\bf x_B}}{\BS x_B} f_{\BS x_A}\cr
{f}_{\bf x_A}^T & 0 & 0\cr
{g}_{\bf x_B}^T + \beta {g}_{{\bf x}_B}^T \PD{f_{{\bf x}_A}}{\bf x_A} & 0 & {g}_{{\bf x}_B} \cdot f_{{\bf x}_A}\cr
\end{pmatrix}
\end{equation}
where $\bf A$ is the following matrix:
\begin {equation}
{\bf A} = \PD{g_{{\bf x}_B}}{{\bf x}_B} + \beta \PD{g_{{\bf x}_B}}{{\bf x}_B} \PD{f_{{\bf x}_A}}{{\bf x}_A}
\end{equation}
Evaluation of ${\bf J}^{-1}$ can be done again by the $LU$ decomposition.


\subsubsection{Relaxed Newton-Raphson method}
Eq. (\ref{Eq:NRiter}) has been derived using a first order expansion of $F({\bf y})$, but if the matrix ${\bf J}$ is singular or nearly-singular in the proximity of the solution, even with a good initial guess the  NR step can become too large and the NR method unstable.
To solve this problem, a damping factor into Eq. (\ref{Eq:NRiter})
can be conveniently introduced:
\begin{equation}
{\bf y}_{i+1} = {\bf y}_i + \xi_i {\bf J}^{-1} {\bf F}({\bf y}_i)
\label{Eq:NRrelax}
\end{equation}
where $0 < \xi_i < 1$ and a suitable $\xi_i$ has to be found in order to stabilize the NR method.
First, consider either the system of equations (\ref{Eq:dist8}) or (\ref{Eq:dist5}); let $d{\bf y}_i = \xi_i \ (d{\bf x}_A,d{\bf x}_B,d\alpha^2,d\beta)$ be the step predicted by NR. The relative variations of $\alpha^2$, $\beta$ can be bounded, imposing the conditions $\xi_i |d\alpha^2| = \xi_i |\alpha d\alpha|= \epsilon_r \alpha^2$ and $\xi_i  |d\beta|  = \epsilon_r  |\beta|$ with $0<\epsilon_r<1$.

 Taking the modulus of both sides of Eq.(\ref{Eq:dist8a}) one gets $\alpha^2  =  \|f_{{\bf x}_A}\| / \| g_{{\bf x}_B} \| $, obtaining the condition:
\begin{equation} 
\xi_i =  \frac{ \epsilon_r \|f_{{\bf x}_A} \|}{2 \|g_{{\bf x}_B}\| |\alpha|}
\label{Eq:xi1a} 
\end{equation}

Similarly, from  Eq. (\ref{Eq:dist8d}), one obtains the equation $\| {\bf x}_A - {\bf x}_B \| = \| f_{{\bf x}_A}\| |\beta|$. Assuming that $\| {\bf x}_A - {\bf x}_B \| \sim l$, where $l$ is a typical distance of the system, provides the condition:
\begin{equation}
\xi_i = \frac{\epsilon_r l}{|\beta| \| f_{{\bf x}_A}\|}
\label{Eq:xi2}
\end{equation}

Therefore, the damping factor $\xi_i$ can be chosen such that:
\begin{equation}
\xi_i = \min\left\{\frac{\epsilon_r l}{|\beta| \| f_{{\bf x}_A}\|},
	      \frac{ \epsilon_r \|f_{{\bf x}_A} \|}{2 \|g_{{\bf x}_B}\| |\alpha|}\right \}
\label{Eq:xidist}
\end{equation}
to ensure that the variations of $\alpha$ and $\beta$ remain sufficiently small.







\subsection{Prediction of the collision time}
In an event-driven (ED) algorithm one needs to predict the time of collision of pairs of HRBs.
This means to evaluate --- given two objects at time $t=0$ and 
the distance as a function of time $d(t)$  ---   
the smallest time $t_c>0$ such that $d(t_c)=0$.
If a good guess of the contact point and time is provided, solving 
a proper set of equations through a NR method (as it will be shown shortly)
allows to find the exact contact point and time.
In the following it will also be shown how to calculate
in a very efficient and simple way such an initial guess,
exploiting the evaluation of the distance between two rigid bodies.

\subsubsection{Newton-Raphson method for the contact time}
The point and time of collision satisfy the following equations:
\begin{subequations} \label{Eq:contime}
\begin{equation} \label{Eq:contimea}
f_{\BS x}(\BS x, t) + \alpha^2 g_{\BS x}(\BS x,t) = 0
\end{equation}
\begin{equation} \label{Eq:contimec}
f(\BS x, t) = 0
\end{equation}
\begin{equation} \label{Eq:contimed}
g(\BS x, t) = 0
\end{equation}
\end{subequations}
where $f$ and $g$ now depend also on time because the objects move. 
Again, it is appropriate to employ the NR method to solve such a system; the Jacobian for the NR 
in this case turns out to be:

\begin{equation}
\BS J = \begin{pmatrix} 
 \PD{f_{\bf x}}{\bf x}+\alpha^2\PD{g_{\bf x}}{\bf x}& 2\alpha g_{\bf x} &\PD{f_{\bf x}}{t}+\alpha^2\PD{g_{\bf x}}{t}\cr
{f}_{\bf x}^T & 0 & f_t\cr
{g}_{\bf x}^T & 0 & g_t
\end{pmatrix}
\label{Eq:jaccontime}
\end{equation}
where $f_t = \PD{f}{t}$ and $g_t = \PD{g}{t}$. As discussed in \cite{Cleary97}, such method becomes very unstable even for simple convex objects unless one starts from a very good initial guess. To construct such a guess, one can find a good bracketing of the collision time,  i.e. two times  $t_1$, $t_2$ such that:
\begin{enumerate}
\item $d(t_1) > 0$ and $d(t_2) < 0$
\item $t_2 - t_1$ is small enough to have at the most one collision within $(t_1,t_2)$ 
\item $d(t_1)$, $d(t_2)$ are small enough in order to avoid any instability of the NR method (see above). 
\end{enumerate}
Once the latter bracketing has been found, the initial time for the NR is set to $t_1$, while a good initial guess for the contact point will be halfway on the distance between the two bodies at $t_1$.

\subsubsection{Bracketing the contact time}\label{sssec:bracket_time}
\begin{figure}
\begin{center}
\includegraphics[width=0.60\linewidth]{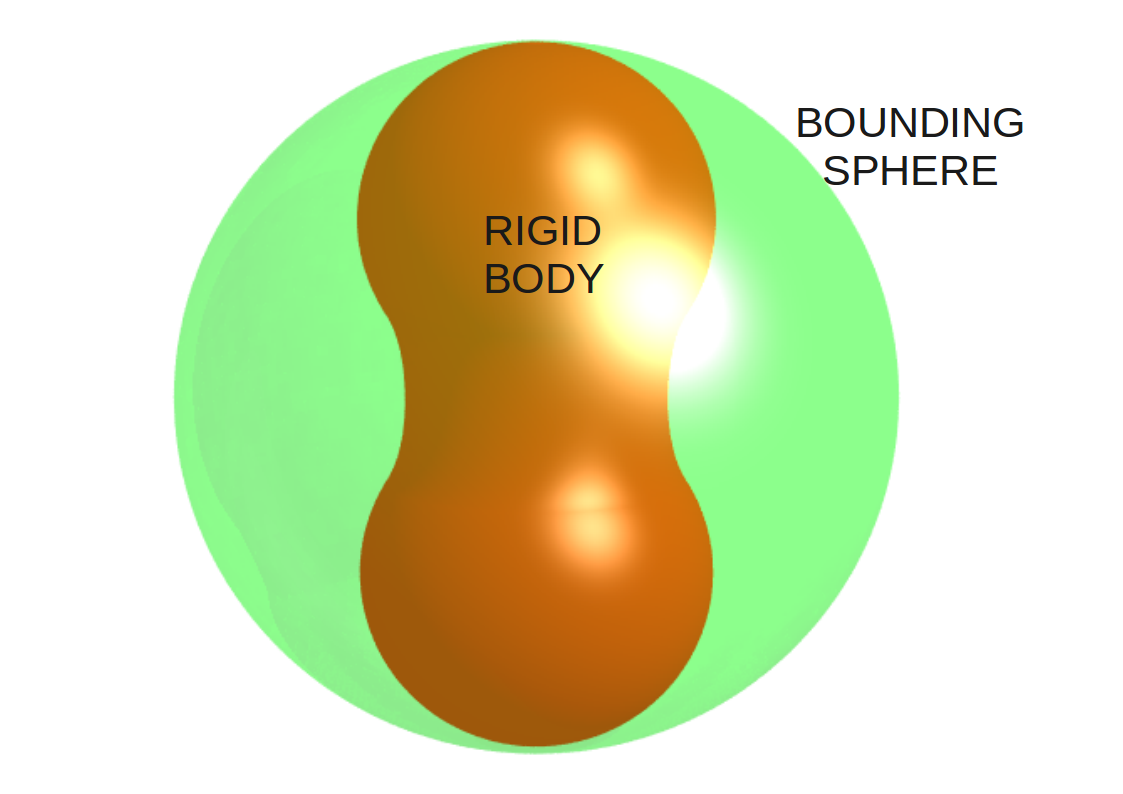}
\caption{\label{fig:centroid} Minimal bounding sphere for a rigid body.}
\end{center}
\end{figure}

To first bracket the contact time a ``bounding sphere" (BS) may be used (Fig.\ref{fig:centroid}).  The BS of a given HRB $A$ with center $\BS r_A$ is the smallest sphere centered at $\BS r_A$ that encloses $A$.
Given two rigid bodies $A$ and $B$ at time $t=0$ and their BSs $C_A$ and $C_B$, one must indeed consider three possible cases:
\begin{enumerate}
\item $C_A$ and $C_B$ do not overlap and they will not collide: in this case $A$ and $B$ won't collide either.
\item $C_A$ and $C_B$ do not overlap but they will collide: in this case one has to search for a collision 
within the time interval $[t_1,t_2]$ where $t_1$ is the time when the two BSs collide and start overlapping and $t_2$, which is greater than $t_1$, is the time when they just cease overlapping.
\item $C_A$ and $C_B$ overlap: in this case one has to search for a collision within the time interval  $[t_1,t_2]$ where $t_1=0$ and $t_2 > t_1$ is the time at which the two BSs 
stop overlapping.
\end{enumerate}
Since the collision prediction between hard spheres (i.e. BSs) is extremely fast this technique improves the performance by reducing the number of collisions to check and provides a first bracketing $(t_1,t_2)$ of the 
collision time.

To further improve this bracketing, the following overestimate of the rate of variation of the distance can be established: 
\begin{equation}
\dot d(t) 
\le \| {\bf v}_A - { \bf v}_B \| +  \| { \bf w}_A \| L_A +
\| { \bf w}_B \| L_B
\label{Eq:distOver}
\end{equation}
where the dot indicates the derivation with respect to time; ${\bf v}_A$ and ${\bf v}_B$ are the velocities 
of the centers of mass of $A$ and $B$; ${\bf w}_A$ and ${\bf w}_B$ are the angular velocities of $A$ and $B$, 
and  the lenghts $L_A$, $L_B$ are such that
\begin{subequations}
\label{Eq:lalb}
\begin{equation}
L_A \ge \max_{{f({\bf r}')\le 0}} \{\|{\bf r}'-{\bf r}_A\|\} 
\end{equation}
\begin{equation}
L_B \ge \max_{{g({\bf r}')\le 0}} \{ \|{\bf r}'-{\bf r}_B\|\} 
\end{equation}
\end{subequations}
where ${ \bf r}_A$, ${\bf r}_B$ are the centers of mass of the two rigid bodies.
To derive Eq.(\ref{Eq:distOver}), consider the case $\dot d(t) > 0$.
In this case, for a positive time increment $dt > 0$ one has:
\begin{equation}
|\dot d(t)| =\frac{\|\BS{x}_A(t+dt) - \BS{x}_B(t+dt)\| - \| \BS{x}_A(t) - \BS{x}_B(t)\|}{dt} 
\label{Eq:timederd}
\end{equation}
where ${\bf x}_A$ and ${\bf x}_B$ are two points belonging to the two rigid bodies such that:
\begin{equation}
d(t) = \| {\bf x}_A(t) - {\bf x}_B(t)\|
\end{equation}
If $\BS{\dot x}_A(t)$ and $\BS{\dot x}_A(t)$ are the velocities at time $t$ of the points $\BS{x}_A(t)$ and $\BS{x}_B(t)$ respectively, it results that:
\begin{equation}
\|\BS{x}_A(t+dt) - \BS{x}_B(t+dt)\| \le  \|\BS{x}_A(t) - \BS{x}_B(t) \| + \| (\BS{\dot x}_A(t) - \BS{\dot x}_B(t)) dt \|
\label{Eq:ddover}
\end{equation}
because of Eq. (\ref{Eq:distdef}), that defines the distance.
From Eq. (\ref{Eq:ddover}) and Eq.(\ref{Eq:timederd}) the following inequality is obtained:
\begin{equation}
|\dot d(t)| \le \| {\bf v}_A - { \bf v}_B \| +  \| { \bf w}_A \times ({\bf x}_A(t)-{\bf r}_A(t)) \| + \|   
{ \bf w}_B \times ({ \bf x}_B(t) - {\bf r}_B(t)) \|
\end{equation}
and Eq. (\ref{Eq:distOver}) results from the fact that:
\begin{subequations}
\label{Eq:reldistover}
\begin{equation}
\| { \bf x}_A(t) - {\bf r}_A(t) \| \le L_A
\end{equation}
\begin{equation}
\| { \bf x}_B(t) - {\bf r}_B(t) \| \le L_B
\end{equation}
\end{subequations}
The case $\dot d(t) < 0$ is similar. Indeed, performing the substitutions $t + dt\rightarrow t'$ and 
$dt \rightarrow -dt'$ into Eq. (\ref{Eq:timederd}), Eq. (\ref{Eq:distOver}) is obtained again.


With such an overestimate of $\dot d(t)$, that from now on will be called $\dot d_{max}$, an efficient strategy for the refinement of the bracketing $(t_1,t_2)$ of the collision time and point is the following one:

\begin{enumerate}
\label{enum:HEalgo}
\item Set $t=t_1$.
\item Evaluate the distance $d(t)$ at time $t$.
\item Choose a time increment $\Delta t$ as follows:
\begin{equation}	
\Delta t =  
\begin{cases}
\frac{d(t)}{\dot d_{max}}, & \hbox{if}\> d(t) > \epsilon_d\cr
\frac{\epsilon_d}{\dot d_{max}}, & \hbox{otherwise.} \cr
\end{cases}
\end{equation}
where $\epsilon_d \ll \min\{L_A,L_B\}$.
\item Evaluate the distance at time $t+\Delta t$.
\item If $d(t+\Delta t) < 0$ and $d(t) > 0$, then $t_1=t$ and $t_2=t+\Delta t$, 
find the collision time and point via NR and terminate. An initial guess for this NR 
may be obtained through a quadratic interpolation of the points $(t,d(t))$, 
$(t+\Delta t / 2,d(t+\Delta t /2)$ and $(t+\Delta t, d(t+\Delta t)$.
\item If both $0 < d(t+\Delta t) < \epsilon_d$ and $0 < d(t) < \epsilon_d$, a ``grazing" collision may occur between $t$ and $t+\Delta t$ (because distance may be first diminishing, then growing). To look for a possible collision, evaluate the distance $d(t+\Delta t /2)$ and perform a quadratic interpolation of the $3$ points ($(t,d(t))$, $(t+\Delta t/2, d(t+\Delta t/2)$, $(t+\Delta t, d(t+\Delta t)$). 
There are $3$ possible cases:
\begin{enumerate}
\item The parabola has a minimum and $d(t_m) < 0$, then set $t_1=t$ and $t_2=t_m$,
find the first zero $t_z$ of this parabola and use $t_z$ and $d(t_z)$ to start a NR to find the collision time and point, and terminate.
\item If the resulting parabola has a minimum $t_m$ between $t$ and $t+\Delta t$,
but $d(t_m) > 0$, then find the minimum $t_{mb}$ of $d(t)$ between  $t$ and $t+\Delta t$
with the maximum possible accuracy (using a $1$-dimensional Brent's method for finding minima).
If $d(t_{mb}) < 0$ then set   $t_1=t$ and $t_2=t_{mb}$ and find the first zero $t_z$ of $d(t)$ 
between $t_1$ and $t_2$ (with a moderate accuracy, because $t_z$ will serve
only as initial guess for the NR).
Use $t_z$ and $d(t_z)$ to start a NR to find the collision time and point and terminate.
\item If the parabola does not have a minimum then keep searching (i.e. go to step~\ref{item:incstep}).
\end{enumerate}
\item \label{item:incstep} Increment time by $\Delta t$, i.e. $t\rightarrow t+\Delta t$.
\item if $t > t_2$ terminate (no collision has been found).
\item Go to step 2.
\end{enumerate}
Note that if starting at time $t$ the two rigid bodies collide at time $t_c > t$ and $d(t) > \epsilon_d$, then our over-estimating procedure enforces $\Delta t < t_c - t$.
If  the chosen $\epsilon_d$ is small enough the distance $d(t)$ has only one minimum within 
the time interval $[t,t+\Delta t]$ and the above scheme for predicting HRBs collisions ensures that ``grazin'' collisions are properly handled within machine accuracy, i.e. all ``grazing" collisions are correctly predicted. 
According to step 6) in the above scheme, indeed, if $d(t) > 0$ and $d(t+\Delta t)>0$, a quadratic interpolation is used
to search for a negative minimum (i.e. a minimum such that $d(t_m) < 0$). If such a negative minimum 
can not be found, an attempt to find it is made with the maximum possible accuracy using a suitable numerical method (e.g. {\it Brent's method}).
We stress that  $\epsilon_d$ must not be chosen with unreasonably tight tolerance not to miss grazing collision within machine accuracy and, although finding the minimum with Brent's method can be time-consuming, this method is only a second 
option after the quadratic interpolation, which is faster and finds the minimum in the majority of cases.
\subsection{Elastic collision of two hard rigid bodies}
As two rigid bodies collide, one has to evaluate the new velocities of centers of mass and the new angular velocities after the collision. Imposing the conservation of impulse, angular momentum and energy, the velocities after the collision can be evaluated as follows:
\begin{subequations} \label{Eq:elcoll}
\begin{equation}
\BS v_A \rightarrow  \BS v_A + m_A^{-1} \Delta p_{AB} \BS{\hat  n}
\label{Eq:elcolla}
\end{equation}
\begin{equation}
\BS v_B \rightarrow  \BS v_B - m_B^{-1} \Delta p_{AB} \BS{\hat  n}
\label{Eq:elcollb}
\end{equation}
\begin{equation}
\BS w_A \rightarrow \BS w_A + \Delta p_{AB} I_A^{-1} (\BS r_A -\BS x_C)\times  \BS{\hat  n}
\label{Eq:elcollc}
\end{equation}
\begin{equation}
\BS w_B \rightarrow \BS w_B - \Delta p_{AB} I_B^{-1} (\BS r_B -\BS x_C)\times  \BS{\hat  n}
\label{Eq:elcolld}
\end{equation}
\end{subequations}
where  $\BS x_C$ is the contact point, 
$\BS{\hat  n}$ is a unit vector perpendicular to both surfaces at $\BS x_C$,
$I_A$, $I_B$ are the moments of inertia of the two colliding rigid bodies, 
$m_A$, $m_B$ are their masses and 
\begin{equation}
\Delta p_{AB} = \frac{2 (\BS v_A + \BS w_A\times\BS (\BS x_C - \BS r_A ) 
	- \BS v_B - \BS w_B\times\BS (\BS x_C - \BS r_B ))}
		{m_A^{-1} + m_B^{-1} + I_A^{-1}\|(\BS r_A -\BS x_C)\times  \BS{\hat  n}\| +
		I_B^{-1}\|(\BS r_B -\BS x_C)\times  \BS{\hat  n}\| } 
\label{Eq:factor}
\end{equation}
\subsection{Linked lists for bounding spheres}
\label{sec:linklists}
Predicting collisions is the most computationally demanding part of an EDMD. In order to speed up a EDMD of hard spheres, one can use linked lists \cite{RapaBook} to avoid to check all the $N^2$ possible collisions among $N$ objects.  In the linked list method, the simulation box is partitioned into $M^3$ cells and only collisions 
between particles inside the same cell or its $26$ adjacent cells are accounted for. This means also that, whenever an object crosses a cell boundary going from cell $a$ to a new cell $b$, 
it has to be removed from cell $a$ and added to cell $b$.

As a first step, one can recover the same method  using the BSs location as particles. In this case,
the cubic box of side $L$ containing $N$ identical rigid bodies is divided into $M^3$ cells so that each cell has side length of the order of the BS diameter $\sigma_c$.
After that linked lists of these BSs are built and updated as in a ordinary EDMD of hard spheres \cite{RapaBook}.
To predict collisions of a rigid body A, one takes into account only the rigid bodies that have their BSs  inside the $26$ adjacent cells or in the same cell as $A$ (see \cite{RapaBook} for more details).

Unfortunately, BSs are not very efficient if the volume of the BS is much bigger than the volume of the rigid body, as it happens for example for ellipsoids of high elongation \cite{Donev05a,Donev05b}. In this case, the number of possible collisions  which must be calculated  makes such procedure 
computationally inefficient. In Section \ref{sec:NNL} an alternative and more efficient method for objects with high elongations will be illustrated.

\subsection{Putting all together: hard rigid bodies event-driven algorithm}
\label{sec:rbED}

In an ED algorithm many events may occur, such as collisions between particles, cell crossing (if one uses linked lists), saving of a system snapshot, output of measured quantities, etc. All these events should be ordered in a calendar so that the next event to happen can be easily retrieved; at the same time, insertion and deletion of events in the calendar 
should be done as quickly as possible.

One elegant approach was introduced almost thirty years ago by Rapaport \cite{RapaBook}, who proposed to arrange all the events into an ordered binary tree (calendar of events), so that insertion, deletion and retrieving of events could be done with an efficiency $O(\log {\cal N})$, $O(1)$ and $O(\log {\cal N})$ 
respectively, where ${\cal N}$ is the number of events in the calendar.  This solution is adopted to handle events calendar in our simulations; all the details of this method can be found in \cite{RapaBook}.

Considering that all the tools to develop a standard ED algorithm for hard rigid bodies have been illustrated, 
the algorithm can be resumed as follows:

\begin{enumerate}
\item Initialize the events calendar (predict collisions, cell-crossings, etc.).
\item Retrieve next event $\cal E$ and set the simulation time to the time of this event. 
\item If final time has been reached, terminate.
\item If $\cal E$ is a collision between particles $A$ and $B$ then:
\begin{enumerate}
\item change angular and center-of-mass velocities of $A$ and $B$ according to Eqs. (\ref{Eq:elcoll})
\item remove from calendar all events (collisions, cell-crossings) in which $A$ and $B$ are involved.
\item predict and schedule all possible collisions for $A$ and $B$.
\item predict and schedule the two cell-crossings events for $A$ and $B$.
\end{enumerate}
\item If $\cal E$ is a cell crossing of a certain rigid body $A$:
\begin{enumerate}
\item update linked lists accordingly.
\item remove from calendar all events (collisions, cell-crossings) in which $A$ is involved.
\item predict and schedule all possible collisions for $A$ using the updated linked lists.
\end{enumerate}
\item go to step 2.
\end{enumerate}

The novel aspect of the present algorithm is in the time-consuming step (4)c, for which we have provided in the previous section  the implementation details. 

\section{Hard ellipsoids with an axis of symmetry}\label{sec:EMD4HE}
In this Section the details about a specific case for which the algorithm 
has been tested will be provided \cite{ourHEstatic}: hard ellipsoids with an axis of symmetry. Such ellipsoids are characterized by the elongation $X_0$, i.e. the ratio of the length of the symmetry axis with respect to any of the other axes.
A new efficient implementation of nearest neighbor lists (NNL) for 
hard ellipsoids will be also discussed and a careful test of the performance of this approach will be shown.

\subsection{Evaluation of the distance between two ellipsoids}
The surface of a hard ellipsoid can be implicitly defined as follows:
\begin{equation}
(\BS x - \BS r_A)^T \BS X_A (\BS x - \BS r_A) = 0
\label{Eq:ellsurf}
\end{equation}
where ${\bf r}_A$ is the position of the center of mass of the ellipsoid and $\BS X_A$ is a $3\times 3$ 
positive definite matrix.
In particular if at time $t=0$ the rigid body reference system coincides to the laboratory 
reference system it turns out that the free evolving hard ellipsoid surface is:
\begin{equation}
X_A = \BS R_A^T(t) \BS X_A(0) \BS R_A(t)
\end{equation}
where 
\begin{equation}
\BS X_A = 
\begin{pmatrix} 
\ a^{-1}\ &\ 0\ &\ 0\ \cr 
0 & b^{-1} & 0\cr 
0 & 0 & c^{-1}
\end{pmatrix}
\end{equation}
and $a,b,c$ are the three semi-axes of the ellipsoids. For hard ellipsoids with an axis of symmetry, the values of two semiaxes are equal.

To evaluate the distance between two ellipsoids $A$ and $B$ at a time $t$ one has to solve 
either Eq. (\ref{Eq:dist8}) or Eq. (\ref{Eq:dist5}) by the Newton-Raphson method.
For Eqs. (\ref{Eq:dist8}) in this particular case the Jacobian becomes:
\begin{equation}
{\bf J} = 
\begin{pmatrix}
2 {\bf X}_A  & 2 \alpha^2 {\bf X}_B  &
2\alpha {\bf n}_B & {\bf 0} \cr
{\bf n}_A^T & {\bf 0}^T  & 0 & 0 \cr
{\bf 0}^T  & {\bf n}_B^T & 0 & 0 \cr
{\bf I} + 2\beta {\bf X}_A & -{\bf I} & 0 & f_{\bf x}\cr
\end{pmatrix}
\end{equation}
where $ \BS n_A = 2 \BS X_A \BS (\BS x_A -  \BS r_A) $ , 
$ \BS n_B = 2 \BS X_B \BS (\BS x_B -  \BS r_B)$
and 
\begin{equation}
\BS X_\mu = \BS R^T_\mu(t) \BS X_\mu(0) \BS R_\mu(t)
\end{equation}
with $\mu \in \{A,B\}$.

For Eqs. (\ref{Eq:dist5}) the Jacobian turns out to be:
\begin{equation}
{\bf J} = 
\begin{pmatrix}
2 {\bf X}^A + \alpha^2 {\bf A} & 2\alpha {\bf n}_B & {\bf c}\cr
{\bf n}_A^T  & 0 & 0 \cr
{\bf n}_B^T + {\bf d}^T & 0 & \BS n_B\cdot\BS n_A\cr
\end{pmatrix}
\end{equation}
where $\BS n_A$, $\BS n_B$ and $\BS X_\alpha$ are the same as before,  $\BS x_B = \BS x_A + \beta\BS n_A$ and
\begin{subequations}
\begin{equation}
\BS c = 2\alpha^2\ \BS X_B\cdot \BS n_A
\end{equation}
\begin{equation}
\BS d = 2\beta\ \BS n_B^T \cdot \BS X_A
\end{equation}
\begin{equation}
\BS A = 2 \alpha^2 \BS X_B \cdot (2 \beta \BS X_A + \BS I) 
\end{equation}
\end{subequations}

\subsection{A better guess of the distance}\label{ssec:HomoGuess}
As already discussed, the NR method needs a good guess of the starting point and 
for this purpose a steepest descent method has been presented in Sec. \ref{sec:iniguess}.
For ellipsoids with moderate elongations ($0.2 < X_0 < 5.0$), the initial guess can 
be calculated in an alternative and very simple way.
The simplest possibility is to use as a first guess for $\BS x_A$ and $\BS x_B$ the intersections of the vector $\BS r_{AB} = \BS r_A - \BS r_B$,  joining the centers of mass of the two ellipsoids, with their surfaces.
This guess is quite rough and a possible improvement can be achieved, as explained in the following. 

First of all the components of 
${\bf r}_{AB}$ in the reference systems of the two ellipsoids are calculated, i.e.
\begin{subequations}
\begin{equation}
\BS v_A' = -\BS R_A\cdot\BS r_{AB}  
\end{equation}
\begin{equation}
\BS v_B' =  \BS R_B\cdot\BS r_{AB} 
\end{equation}
\end{subequations}
then these two vectors will be scaled by the semi-axes of the ellipsoids
and their components will be transformed back to the laboratory reference system, i.e.:
\begin{subequations}
\begin{equation}
\BS v_A = \BS R_A^T(\BS S \cdot \BS v_A')  
\end{equation}
\begin{equation}
\BS v_B = \BS R_B^T(\BS S \cdot \BS v_B')  
\end{equation}
\end{subequations}
where $\BS R_A$ and $\BS R_B$ are the orientational matrices of $A$ and $B$ respectively and
\begin{equation}
\BS S = \frac{1}{\sqrt{a^2+b^2+c^2}}
\begin{pmatrix}
\ a\ &\ 0\ &\ 0\ \cr
0 & b & 0\cr
0 & 0 & c\cr
\end{pmatrix}
\end{equation}
Finally the intersections with the HE surfaces of the vector ${\BS{v}}_A$ 
centered at the center of mass of $A$ and 
${\BS{v}}_B$ centered at the center of mass of $B$ are computed and 
these two points are 
the desired guesses for $\BS x_A$ and $\BS x_B$ for the NR. 
The effects of such trasformations are shown in the right panel of Fig.(\ref{fig:ABguess}).

At this point, distance evaluation can be started with the initial values for $\alpha$ and $\beta$ set to $0$. 
This method provides a speed-up of around $10\%$.

\subsection{Nearest Neighbour Lists}\label{sec:NNL}
\begin{figure}
\begin{center}
\includegraphics[width=0.75\linewidth]{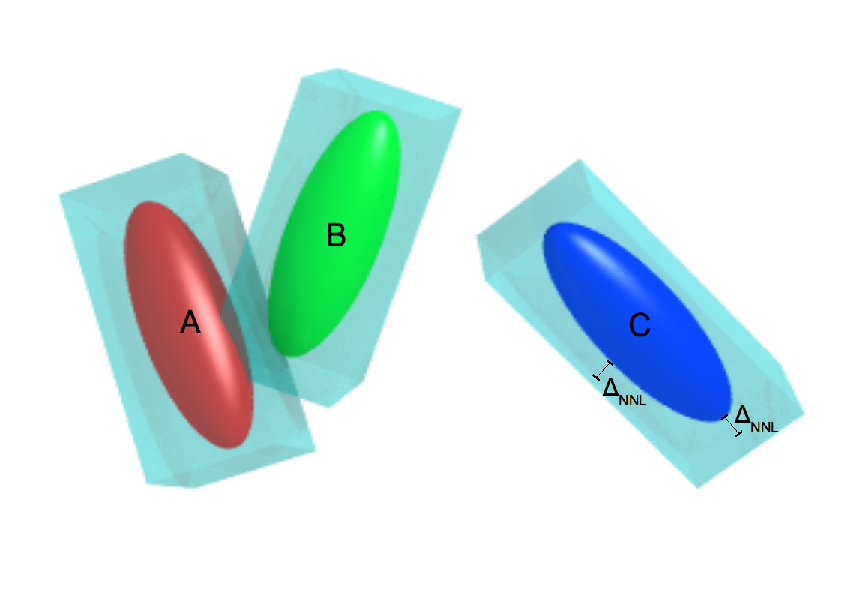}
\vspace{-1cm}
\caption{\label{fig:BoundingBox} Nearest neighbour lists are built through OBPs:  ellipsoids A and B are nearest neighbors, and they may collide among them before colliding with their OBP; neither A or B can collide with C during this time. The "thickness" of the OBP is $\Delta_{NNL}$.}
\end{center}
\end{figure}

Increasing the elongation of the ellipsoids, the linked list method illustrated before becomes progressively less efficient at moderate and high densities  \cite{Donev05a}.
Indeed given one ellipsoid $A$, when using the linked lists one has to
predict the collision times of $A$ with all ellipsoids in the cell of $A$ and with all
ellipsoid in the $26$ adjacent cells. In the case of rotationally symmetric ellipsoids,
the number of collision time predictions grows as $X_0^2$ if $X_0 > 1$
and as $1/X_0$ if $X_0 < 1$ \cite{Donev05b} (see also Sec. \ref{Sec:elongdep}).
In order to reduce the number of predictions at very high or very small elongations, Donev {\it et al.} \cite{Donev05a} suggest 
to surround each particle $A$ with a {\it bounding neighborhood} having the same shape as $A$ (i.e. in the case of HEs they use
ellipsoids) and to predict collisions only between particles having overlapping bounding neighborhoods.
Similarly to what is proposed by Donev {\it et al.} \cite{Donev05a} we suggest  to build an oriented bounding parallelepiped (OBP),
instead of an ellipsoid, at a certain time $t$ around each HE and to predict collisions only between HEs having overlapping parallelepipeds (Fig.~\ref{fig:BoundingBox}).

Each parallellepiped encloses the corresponding ellipsoid completely, and it is centered at the center of mass of the ellispoid itself. More precisely, given an ellipsoid $A$ with semi-axes $a,b,c$ and center of mass $\BS r_A$, the vertices $\BS v_\alpha$ of the parallelepiped, with $\alpha\in\{1\ldots 6\}$ are:
\begin{equation}
\BS v_\alpha = \BS r_A + \sigma_2(\alpha) (s_{\sigma_1(\alpha)} + \Delta_{NNL}) \BS u_{\sigma_1(\alpha)}   
\end{equation}
where 
\begin{eqnarray}
\sigma_1(\alpha) &=& 1 + (\alpha-1) \div 2\cr
\sigma_2(\alpha) &=& 2\,(\alpha\bmod 2) - 1
\end{eqnarray}
and $\BS s=(s_1,s_2,s_3) = (a,b,c)$.
Moreover $\{\BS u_i\}_{i\in\{1,2,3\}}$ are the principal axes of the ellipsoid, i.e. they are such that:
\begin{equation}
\BS X_A \BS u_\alpha = s_\alpha\BS u_\alpha 
\end{equation}
and $\Delta_{NNL}$ is a positive parameter that can be tuned to optimize the performance 
of the NNL (see Sec. \ref{sec:nnlopt}).

Given an ellipsoid $A$, the set of ellipsoids having their parallelepipeds overlapping 
with the parallelepiped enclosing $A$, is the NNL of $A$.
Each parallelepiped $i$ is immobile and contains only the $i$-th ellipsoid; if $t_i$ is the time when the ellipsoid will collide with its containing parallelepiped, its NNL will have to be rebuilt not after the time $t_{r} = \min_i \{ t_i \}$.

In addition, if a collision between two ellipsoids $i$ and $j$ occurs, then the new contact times $t_i$  and $t_j$ of these two ellipsoids with their parallelepipeds will have to be evaluated and the new time for rebuilding the NNL lists, i.e. $t_{r}' = \min \{ t_r, t_i, t_j \}$, will have be set.
 
\subsubsection{Distance between a rigid body and a plane}
\begin{figure}
\begin{center}
\includegraphics[width=0.65\linewidth]{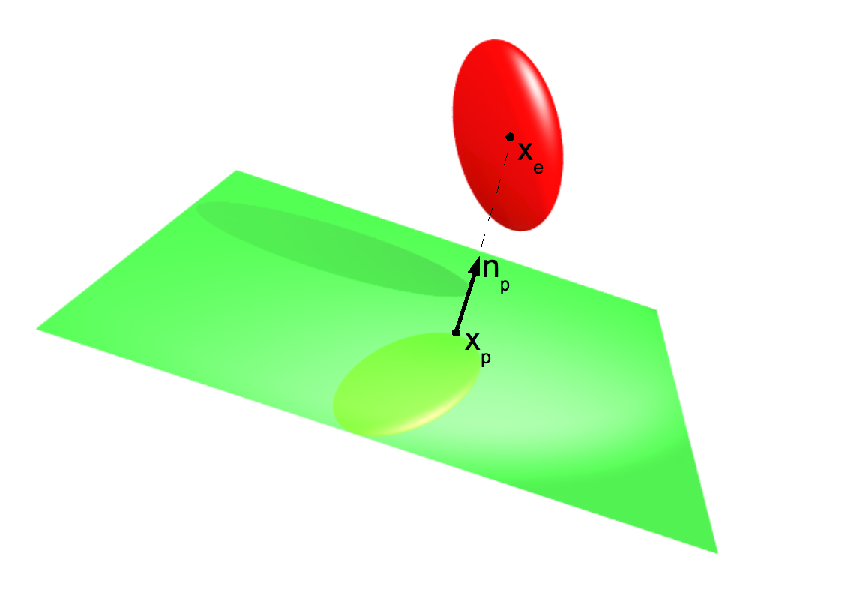}
\vspace{-0.8cm}
\caption{\label{fig:ABguess_plane} The initial NR guess to calculate the distance between a plane $P$ and an ellipsoid $E$ uses the points ${\bf x}_E$ and   ${\bf x}_P$. Here ${\bf n}_P$ is the normal to the plane; the center-of-mass of the ellipsoid is on the line from ${\bf x}_P$ with direction ${\bf n}_P$.}
\end{center}
\end{figure}

For predicting the time of collision between an ellipsoid and a parallelepiped the distance between an ellipsoid and each of the $6$ faces of the  parallelepiped must be calculated.
This means that one has to evaluate the distance between the surfaces defined through Eqs. (\ref{Eq:ellsurf}) and a plane defined through the following equation:
\begin{equation}
\BS n_P \cdot ( \BS x - \BS r_P) = 0 
\label{Eq:plane}
\end{equation}
where both $\BS r _P$ and $\BS n_P$ do not depend on time because the plane is immobile (Fig. \ref{fig:ABguess_plane}).

Again a NR method can be used to calculate this distance.
In Sec. 2 the second rigid body $B$ defined by Eq. (\ref{Eq:shapeg}) may also be a plane 
(as defined through Eq. (\ref{Eq:plane})) or, alternatively,
a plane can be thought as a limiting case of a very large ellipsoids,
in which case the Jacobian needed to solve Eqs. (\ref{Eq:dist8}) turns out to be:
\begin{equation}
\label{eq:distHEplane}
\BS J = 
\begin{pmatrix}
2 {\bf X}_A& {\bf 0}  & -2 \alpha {\bf n}_P & {\bf 0} \cr
{\bf n}_A^T & {\bf 0}^T & 0 & 0 \cr
{\bf 0}^T & {\bf n}_P^T & 0 & 0\cr
{\bf I} & -{\bf I} & 0 & {\bf n}_P \cr
\end{pmatrix}
\end{equation}

while the Jacobian for solving Eqs. (\ref{Eq:dist5}) is:
\begin{equation}
\BS J=
\begin{pmatrix}
2 {\bf X}_A  & -2 \alpha {\bf n}_P & {\bf 0} \cr
{\bf n}_A^T  & 0 & 0 \cr
{\bf n}_P^T & 0 & \BS n_P \cdot \BS n_P\cr
\end{pmatrix}
\end{equation}
A good guess is required also to solve Eqs. (\ref{Eq:dist8}) or Eqs. (\ref{Eq:dist5}) through a NR method.
Let $l$ be the axis passing through the center of the ellipsoid and parallel
to the vector $\BS n_P$.
If intersections points $\BS x_E$ and $\BS x_P$ of $l$ with the ellipsoid  and the plane respectively are evaluated, then these two points (if Eqs. (\ref{Eq:dist8}) are used) or only $\BS x_E$ (if Eqs. (\ref{Eq:dist5}) are used)
proved to be a good initial guess for the NR method.

\subsubsection{Prediction of the collision time}

Given a good bracketing of collision time of an ellipsoid $A$ with one
of the six planes, Eqs. (\ref{Eq:contime}), where $f$ defines 
the surface of an ellipsoid and $g$ the surface of a plane, can be solved numerically.
The Jacobian to use in the NR is the following:
\begin{equation}
\label{eq:colltimeHEplane}
\BS J= 
\begin{pmatrix}
2 {\bf X}_A & -2\alpha {\bf n}_P & {\bf h}\cr
{\bf n}_A^T  & 0 & k\cr
{\bf n}_P^T & 0 & 0\cr
\end{pmatrix}
\end{equation}
where  
$ \BS h = \tilde\Omega\; \tilde{\BS x} - \BS X_A\BS v_A $,
$k = - {\BS v}_A^T \BS X_A \tilde{\BS x} + \tilde{\BS x}^T \;\tilde\Omega\; \tilde{\BS x} - \tilde{\BS x}\; \BS X_A \BS v_A$ 
and $ \tilde\Omega =\BS\Omega_A\BS X_A - \BS X_A\BS\Omega_A$, 
$\tilde{\BS x} =\BS x - \BS r_A$,.
Here $\BS\Omega_A$ is the matrix $\BS\Omega$ defined in Eq. (\ref{Eq:omegmat}) corresponding to the  angular velocity $\BS w_A$ of the ellipsoid $A$ and ${\bf v}_A$ is the center of mass velocity of $A$.

As for the collision of two ellipsoids, the collision time can be bracketed  
using an overestimate of $\dot d(t)$ 
(being $d(t)$ the distance between an ellipsoid and a plane of the parallelepiped).
If the ellipsoid surface is the locus of points such that $f(\BS x)=0$ and the plane is defined as in Eq. (\ref{Eq:plane}), the distance between an ellipsoid and a plane can be defined as follows:
\begin{equation}
d = \min_{f(\BS x_A)=0} | \BS x_A \cdot {\hat{\BS n}}_P| 
\label{Eq:distellplane}
\end{equation}
where ${\hat{\BS n}}_P = \BS n_P / \|\BS n_P\|$. 
Similarly to what has been done for two ellipsoids, it can be proved that
\begin{equation}
\dot d(t) \le  | \BS v_A\cdot  {\hat{\BS n}}_P| + L_A \| \BS w_A \|
\label{Eq:distplover}
\end{equation}
From now on, the latter overestimate will be called $\dot d_{pm}$.
It is now possible to illustrate the algorithm to find the collision time of an ellipsoid with its OBP.
Consider an ellipsoid $A$ and the six planes of the corresponding parallelepiped labelled by 
$\{p_i\}_{1 < i < 6}$.
With respect to each plane there are six different overestimations of $\dot d(t)$, labeled as $d_{pm}^i$, and six different distances $d_i(t)$ at time $t$.
The algorithm is the following one:
\begin{enumerate}
\item set $t=t_1$.
\item Evaluate the distances $d_i(t)$ at time $t$.
\item Choose a time increment $\Delta t$ as follows:
\begin{equation}	
\Delta t =  
\begin{cases}
\min_i\{d_i(t)/\dot d_{pm}^i\}, & \hbox{if}\> d_i(t) > \epsilon_d^{nnl} \hbox{;} \cr
\frac{\epsilon_d^{nnl}}{\max_i\{\dot d_{pm}^i\}}, & \hbox{otherwise.} \cr
\end{cases}
\end{equation}
where $\epsilon_d^{nnl} \ll L_A$.
\item Evaluate the distance at time $t+\Delta t$.
\item If for at least one $i$ one has $d_i(t+\Delta t) < 0$ and $d_i(t) > 0$, then the solution is bracketed. Find the collision time and the collision point through the NR.
For the initial guess of the collision time, use the first zero of the parabola resulting from a quadratic interpolation of points $(t,d(t))$, $(t+\Delta t/2, d(t+\Delta t/2))$ and 
$(t+\Delta t, d(t+\Delta t))$.  After that, terminate.
\item For all distances such that $0 < d_i(t+\Delta t) < \epsilon_d$ and $0 < d_i(t) < \epsilon_d$, evaluate the distance $d_i(t+\Delta t /2)$, perform a quadratic interpolation of these  $3$ points ( $(t,d_i(t))$, $(t+\Delta t/2, d_i(t+\Delta t/2)$, $(t+\Delta t, d_i(t+\Delta t)$) and find if the resulting parabola has any zeros.
If this is the case, refine the position of the smallest zero using again the NR.
\item Increment time by $\Delta t$, i.e. $t\rightarrow t+\Delta t$.
\item if time is greater than $t_2$, terminate (no collisions have been found).
\item Go to step 2.
\end{enumerate}
Grazing collisions in this specific case are dealt with differently from collisions between
two HEs. 
The trick is to consider, for predicting the escape time of a HE, a smaller bounding box.
In particular the chosen distance between each of the $6$ planes forming the bounding box 
and the embedded HE is $\Delta_{NNL}-\epsilon_d^{nnl}$.
With this choice of the bounding box dimensions, if a collision is missed, due to 
a grazing collision, the HE is not outside its bounding box anyway and 
values for $\epsilon_d^{NNL}$ around $10^{-2}\, b$ can be safely chosen.

The times $t_1$ and $t_2$ are evaluated making use of the BS of $A$ as illustrated in the following.
Consider the ellipsoid $A$ and its BS $C_A$ at a certain time $t$ when the NNL has to be rebuilt. Two possible different cases may occur:
\begin{enumerate}
\item $C_A$ is enclosed into the parallelepiped: in this case $t_1$ is the time when 
the BS will collide with one of the six planes of the parallelepiped and $t_2 > t_1$ is the  time when the BS stops intersecting the plane.
\item $C_A$ intersects the parallelepiped: in this case $t_1=t$ and $t_2 > t_1$ is the time at which the two BSs cease to overlap.
\end{enumerate}

\subsubsection{Overlap of Parallelepipeds}
For building NNL all the parallelepipeds overlapping with a given one have to be found. This task can be accomplished quite efficiently, using a technique well known in computer graphics \cite{OBBcolldetect}.
Consider two parallelepipeds corresponding to ellipsoids $A$ and $B$ 
with centers $\BS r_A$ and $\BS r_B$, principal axes $\BS u^A_\alpha$
and $\BS u^B_\alpha$ and semi-axes $a,b,c$.

Consider the following straight lines originating from the center
of $A$:  
\begin{equation}
\BS t_j = \BS r_A + \xi \BS w_j
\end{equation}
with
\begin{equation}
\BS w_j \in \{ \{\BS u^A_\alpha\}_\alpha, \{ \BS u^B_\alpha\}_\alpha, 
\{\BS u^A_\alpha \times \BS u^B_\beta\}_{\alpha,\beta} \}  
\end{equation}
where $\alpha,\beta = 1,2,3$, $\xi\in\Re$ and $j=1\ldots15$ labels all the possible
directions.

The centers of the ellipsoids $\BS r_A$ and $\BS r_B$ will be projected onto all these lines,
obtaining the points $\tilde{\BS r}^A_j$ and $\tilde{\BS r}^B_j$ and the vertices of the two parallelepipeds will be projected as well, obtaining the points $\BS p^A_{i,j}$ and $\BS p^B_{i,j}$, where $i=1\ldots 8$ labels all the possible vertices of a parallelepiped.

Using the projected vertices one can build for each direction $j$ two spheres 
with centers $\tilde{\BS r}^A_j$ and $\tilde{\BS r}^B_j$ and radii:
\begin{eqnarray}
\rho^A_j &=& \max_{i} \{ \|\BS p^A_{i,j} - \tilde{\BS r}^A_j\|\} \cr
\rho^B_j &=& \max_{i} \{ \|\BS p^B_{i,j} - \tilde{\BS r}^B_j\|\}  
\end{eqnarray}
The two parallelepipeds do not overlap if and only if all these pairs of spheres do not overlap.

\subsection{Prediction of the collision time between two ellipsoids}
Exploiting linked lists and BSs, one obtains a time interval $[t_1,t_2]$ that brackets the possible contact time between two ellipsoids $A$ and $B$.
Making also use of NNL, a better bracketing of the collision time 
can be estimated, with the bracketing time interval 
becoming $[t_1,\tilde t_2]$, with: 
\begin{equation}
\tilde t_2 = \min \{ t_2, t_r \}
\end{equation}
where $t_{r}$ is the time at which the NNL rebuild is scheduled.  
Then starting from $t=t_1$ the collision time can be finely bracketed by applying 
the algorithm illustrated in Sec. \ref{sec:rbED} for the general case of hard rigid bodies 
with the substitution $t_2 \rightarrow \tilde t_2$.

The bracketing of the contact point and time for the collision between two HEs
can be used in the NR to solve Eqs. (\ref{Eq:contime}) with the following Jacobian:
\begin{equation}
\BS J =
\begin{pmatrix}
2 ({\bf X}_A + \alpha^2 {\bf X}_B)  &
-2\alpha {\bf n}_B & {\bf h}\cr
{\bf n}_A^T & 0 & k_A\cr
{\bf n}_B^T & 0 & k_B\cr
\end{pmatrix}
\end{equation}
where
\begin{equation}
\BS h = \tilde\Omega_A\; \tilde{\BS x}_A - \BS x_A\BS v_A + \alpha^2 (\tilde\Omega_B\; \tilde{\BS x}_B - \BS x_B\BS v_B)
\end{equation}
and 
\begin{eqnarray}
k_A &=& -{\BS v}_A^T \BS X_A \tilde{\BS x}_A + \tilde{\BS x}_A^T \;\tilde\Omega_A\; \tilde{\BS x}_A - \tilde{\BS x}_A\; \BS X_A \BS v_A\cr
k_B &=& -{\BS v}_B^T \BS X_B \tilde{\BS x}_A + \tilde{\BS x}_A^T \;\tilde\Omega_B\; \tilde{\BS x}_B - \tilde{\BS x}_B\; \BS X_B \BS v_B
\end{eqnarray}
Here, one has:
$\tilde\Omega_A =\BS\Omega_A\BS X_A - \BS X_A\BS\Omega_A$, $\tilde\Omega_B =\BS\Omega_B\BS X_B - \BS X_B\BS\Omega_B$,
$\tilde{\BS x}_A =\BS x - \BS r_A$,
$\tilde{\BS x}_B =\BS x - \BS r_B$,
where $\BS\Omega_A$ is the matrix $\BS\Omega$ defined in Eq. (\ref{Eq:omegmat}) corresponding to the angular velocity $\BS w_A$ of the ellipsoid $A$,  and $\BS\Omega_B$ is the same matrix for the ellipsoid $B$.

\subsection{Event-driven molecular dynamics of Hard Ellipsoids}
The ED molecular dynamics of HEs is similar to what illustrated in Sec. \ref{sec:rbED}.
The only addition is the use of NNL to speed the simulation up.
In the following the scheme of an EDMD of HEs is given:
\begin{enumerate}
\item Initialize the events calendar (predict collisions, cells crossings, etc.).
\item Retrieve next event $\cal E$ and set the simulation time to the time of this event. 
\item If final time has been reached, terminate.
\item If $\cal E$ is the ``NNL rebuild'' event then:
\begin{enumerate}
\item remove all events from calendar.
\item update the system to current time.
\item evaluate $t_r$.
\item check for overlaps between parallelepipeds and build the NNL accordingly.
\item predict all cell-crossings and schedule them.
\item predict all collisions between ellipsoids and schedule them.
\item schedule next NNL rebuild.
\item schedule all other remaining events removed from calendar (output of summary, etc.).
\end{enumerate}
\item If $\cal E$ is a collision between particles $A$ and $B$ then:
\begin{enumerate}
\item change angular and center-of-mass velocities of $A$ and $B$ 
      according to Eqs. (\ref{Eq:elcoll})
\item remove from calendar all events (collisions, cell-crossings) in which $A$ and $B$ 
      are involved.
\item evaluate new collision times $t_A$ and $t_B$ of $A$ and $B$ with their parallelepipeds.
      and schedule a new ``NNL rebuild`` event if $t_A$ or $t_B$ are less than $t_r$.
\item predict and schedule the two cell crossings events for $A$ and $B$.
\item predict and schedule all possible collisions for $A$ and $B$ using the NNL of $A$ and $B$.
\end{enumerate}
\item If $\cal E$ is a cell crossing of a certain rigid body $A$:
\begin{enumerate}
\item update linked lists accordingly.
\item remove from calendar all events (collisions, cell-crossings) in which $A$ is involved.
\item predict and schedule all possible collisions for $A$ using the NNL of $A$.
\end{enumerate}
\item Go back to step 2.
\end{enumerate}
Note that linked lists are used just to rebuild the NNL, that is given a parallelepiped corresponding to an ellipsoid $A$, one searches for overlapping parallelepipeds among all the ones that are in the same cell as $A$ or in one of the $26$ neighbors cells, assuming that the cells side is greater than the length of the diagonal of the parallelepipeds.
To rebuild NNLs Donev {\it et al.} \cite{Donev05b} use, in addition to LL, also a collection of small spheres, called
{\it bounding sphere complex} and in their implementation this method ensures that the cost of building the NNLs
can be controlled increasing the elongation. In our implementation of NNL, that relies on the use of bounding parallelepipeds, 
the average collision time (see discussion in Section \ref{Sec:elongdep})  is quite independent from elongation for $X_0$ up to 
$10$. This is mainly due to the fact that the time needed to check overlaps between bounding parallelepipeds is negligible. 

In the above scheme NNL update is a global event (complete update), 
in  \cite{Donev05b} a method is described that allows to update only NNLs of affected particles after a collision (partial update) .
In their implementation NNL are built using HEs instead of parallelepipeds 
and partial update method outperforms the complete update method (see \cite{Donev05b}). 
Differently in our implementation the time needed to update NNLs at moderate and high densities, 
which is what we are interested in the most, is negligible (see next Section) and the NNLs partial
update method cannot offer any significant speedup.

\section{Performance results.}\label{sec:perform}

To analyze the performance of the algorithm, monodisperse uniaxial HEs in the isotropic liquid phase have been simulated, characterized by the aspect ratio $X_{0}=a/b$ (where $a$ is the length of the revolution axis, $b$ of the other ones) and by the packing fraction $\phi=\pi X_{0}b^{3}\rho/6$ (where $\rho=N/V$
is the number density). $N=256$ particles have been simulated at various volumes $V$ and elongations $X_0$ in a cubic box with periodic boundary conditions.  
The length of the smallest semi-axis is chosen as unit of distance
 and the mass $m$ of the particle as unit of mass ($m=1$). Temperature is  one in units of the Boltzmann constant $k_{B}$; the corresponding unit of time is $\sqrt{ml^{2}/(k_{B}T)}$. 
A spherically symmetric moment of inertia, i.e. $I_{x}=I_{y}=I_{z}=2\, mr^{2}/5$ is chosen,
where  $r=\min(a,b)/2$ is the radius of the sphere inscribed in the ellipsoid. Notice that the value of the moment of inertia along the symmetry axis is not relevant for the present system, since angular velocity around the symmetry axis is zero. Although ellipsoids of revolution behave as symmetric tops, the colliding surfaces will be treated as perfectly smooth (see Eq. (\ref{Eq:elcoll})) and therefore the component of angular velocity along the symmetry axis will be conserved in each collision.
In the following we  indicate as ``reduced time" the simulation time expressed in reduced units,
while ``CPU time" means the real time spent by the  CPU for calculations,
expressed in seconds.  To test the algorithm speed, we use the CPU time per ellipsoid collision, labelled with $\tau_c$ and defined
as:
\begin{equation}
\tau_c = \frac{T_{tot}}{N_{coll}} 
\label{Eq:colltimeperHE}
\end{equation}
where $T_{tot}$ is the (real) time needed to perform $N_{coll}$ collisions during a simulation.

Predicting a collision for a pair  of colliding HEs (labelled by $A$ and $B$) requires the CPU time $\Delta t_{coll}$:
\begin{equation}
\Delta t_{coll} = \sum_l^{N_{pc}^A} \delta t_l^{A} + \sum_l^{N_{pc}^B} \delta t_l^{B} 
\label{Eq:timepercoll}
\end{equation} 
where $N_{pc}^{X}$ indicates the number of
 collisions that have to be predicted for ellipsoid $X$ ($X=A,B$) and with 
 $\delta t_l^{X}$  the CPU time requested to predict one collision. Indeed, 
 after a collision between two HEs, all the possible future events involving the 
two colliding HEs must be predicted.  
If the quantity ``average prediction time" is defined as follows:
\begin{equation}
\langle \delta t \rangle_X \equiv \frac{1}{N_{pc}^{X}} \sum_l^{N_{pc}^X} \delta t_l^{X}
\end{equation} 
where again $X=A,B$.
Eq. (\ref{Eq:timepercoll}) can be rewritten as:
\begin{equation}
\Delta t_{coll} = N_{pc}^A \langle \delta t \rangle_A + N_{pc}^B \langle \delta t \rangle_B
\label{Eq:timepercollSimpl}
\end{equation}  
On passing note that, using LL and/or NNL, $N_{pc}^A$ and $N_{pc}^B$
are $O(1)$ varying the total number $N$ of HEs in the simulation.
Such a number will be generally greater for LL than for NNL, 
but we will discuss this point more accurately later.
Also note that  $\langle \delta t \rangle_A$ and  $\langle \delta t \rangle_B$  depend mostly on the 
average number of Newton-Raphson steps performed to predict the collision time. 
\begin{figure}
\begin{center}
\includegraphics[width=0.6\linewidth]{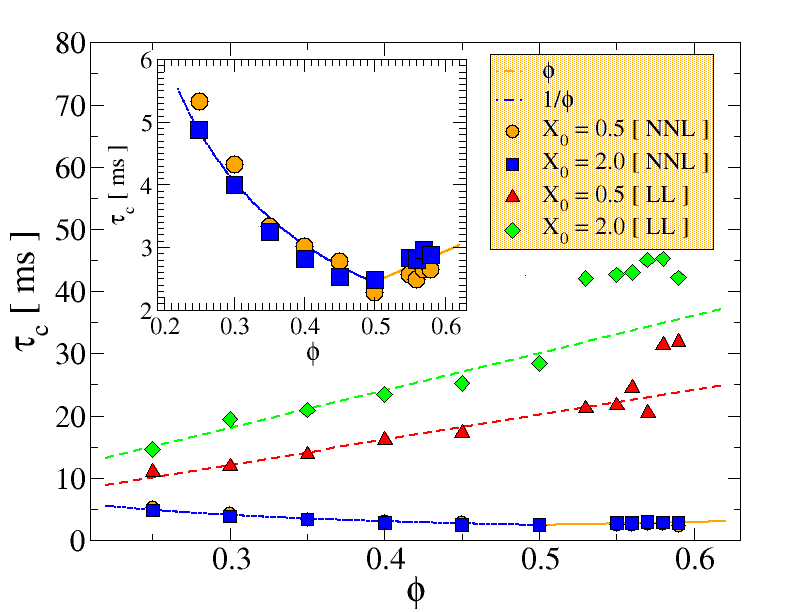}
\caption{\label{fig:CollTimePhi} Time per collision $\tau_c$ in the prolate 
case ($X_0=0.50$) using only LLs (green diamonds) and also NNLs (orange circles) and in the oblate case 
($X_0=2.00$) using only LLs (red triangles) and also NNLs (blue squares).
Dot-dashed lines are guides to the eyes, showing the $\phi$ and $1/\phi$ behavior for the NNLs case, while
dashed lines are guide to the eyes showing a linear behaviour in $\phi$ for the LLs case 
(see discussion in Section \ref{sec:deponphi}).
NNLs show much less density dependence than LLs. 
Note that in all simulations the number of LL cells into which the box is partitioned is not kept fixed.
As a consequence $\tau_c$ vs $\phi$ shows a simple linear dependence up to $\phi\approx 0.55$, where,
due to increasing density, the number of LL cells decreases by one unit breaking the linear
behavior of $\tau_c(\phi)$.
The inset is a zoom of $\tau_c$ vs $\phi$ employing also NNLs.}
\end{center}
\end{figure}
In addition, during the simulation NNL and LL have to be updated and this will 
cost an extra time $\Delta t_{u}$ per HE collision, so that 
the time per ellipsoid collision $\tau_c$ is:
\begin{equation}
\tau_c = \Delta t_{coll} + \Delta t_{u}
\label{Eq:timepercollFull}
\end{equation}  

The contribution of the LLs update  to $\Delta t_{u}$ compared to the NNLs update contribution is always negligible, meaning that
the main contribution comes from the update of the NNLs, because it is more time consuming.
Indeed, NNLs update requires the evaluation of the escape time of each HE from its
bounding box and this is time consuming.
Moreover, at high densities, $\Delta t_{u}$ is usually negligible with respect to $\Delta t_{coll}$,
because the diffusivity of the particles is low at high densities and the average escape time of HE from their bounding boxes becomes quite long.
All simulations are performed on a Intel Xeon CPU 3.06 GHz (codenamed ``Prestonia'') with a L2 cache size of 512 KB 
and a 533 Mhz FSB.

\subsection{Generation  of the initial configurations}
To create the starting configuration at a desired $\phi$, an extremely diluted $\alpha\hbox{-}FCC$ crystal has been melted
\cite{AllTildBook}; afterwards, the particles have been grown independently up to the desired packing fraction (quench in $\phi$ at fixed 
$N$, $X_{0}$).
The details of the growth algorithm will be illustrated in a future publication.
To generate history-independent configurations, tests have been performed on equilibrium configurations. 
To  test the equilibration of the sample, the decay of the self correlation function $F_{self}(q,t) = \frac{1}{N} \sum_i^N e^{ {\bf q}\cdot ({\bf r}_i(t) - {\bf r}_i(0))}$ and  of  the orientational correlation function,$C_2(t) = P_2(\cos\theta(t)) $ - where $P_2(x) = (3 x^2 - 1) / 2$ and $\theta(t)$ is the angle between the symmetry axis at time $t$ and at time zero, have been studied. 

\subsection{Dependence of the algorithm speed on density.}
\label{sec:deponphi}
Elongations $X_0=0.50$ and $2.00$ have been considered, fixing  $\epsilon_d=10^{-5}$ and the thickness of the neighbour shell $\Delta_{NNL}=0.15$.

Using only LLs both in the prolate  and in the oblate 
case (see Fig. \ref{fig:CollTimePhi})), $\tau_c$ increases going from lower to higher densities. 
In this case $\Delta t_u$ is negligible and the average time to predict a collision $\langle \delta t\rangle$ is roughly proportional to the number of HEs inside each LL cell, that in turn is proportional to the volume fraction $\phi$ of the system.


When using both LLs and NNLs the scenario is quite different. 
In Fig.~\ref{fig:CollTimePhi} results from simulations using NNLs are shown, and
it is apparent that there are two different regimes of $\tau_c(\phi)$ at low and high volume fractions $\phi$.
The interpretation of this observed behavior is quite straightforward. Indeed, considering 
Eq.~(\ref{Eq:timepercollFull}),  below $\phi\approx 0.5$ there are few HEs within each NNL, i.e. $N_{pc}^A \approx 0$ and $N_{pc}^B \approx 0$  and the time per collision 
$\tau_c$ is dominated by  $\Delta t_{u}$, that is by the time per collision needed to 
update the NNLs.
In this case $\Delta t_u$ is proportional to the number of average evaluations of the contact 
time of a HE with its bounding box $t_r$, i.e. if $l\propto 1/\phi$ is roughly the mean free path between two HEs collisions:
\begin{equation}
\Delta t_u \propto l / \Delta_{NNL} \propto \frac{1}{\phi}  
\end{equation} 
In Fig.~\ref{fig:CollTimePhi} the $\phi^{-1}$ dependence is evident at small $\phi$.

In contrast, at high densities $\Delta t_{u}$ is negligible and the CPU time needed to predict a collision is dominated by $\Delta t_{coll}$, that in turn depends linearly on 
the number of neighbors within the bounding box of a certain HE, i.e.
\begin{equation}
\tau_c \propto N_A \propto \frac{4\pi}{3}\phi\, (a+\Delta_{NNL}) (b+\Delta_{NNL}) (c+\Delta_{NNL}) - 1  
\end{equation} 
At high volume fractions this equation simplifies to $\tau_c\propto\phi$
and again Fig.~\ref{fig:CollTimePhi} confirms this dependence on $\phi$ at high 
volume fractions.




Finally, comparing NNLs and LLs results in Fig. \ref{fig:CollTimePhi}, 
it is apparent that using NNLs the time per HE collision is significantly smaller  than 
in the case where only LLs are used. Moreover, it 
is far less density-dependent.
This stems from the fact that in Eq. (\ref{Eq:timepercollSimpl})  
$N_{pc}^A$ and $N_{pc}^B$ using NNLs are much smaller 
than the same quantities using LLs (see also discussion in the next section).

%

\subsection{Dependence of algorithm speed on elongation}
\label{Sec:elongdep}
\begin{figure}
\begin{center}
\includegraphics[width=0.49\linewidth]{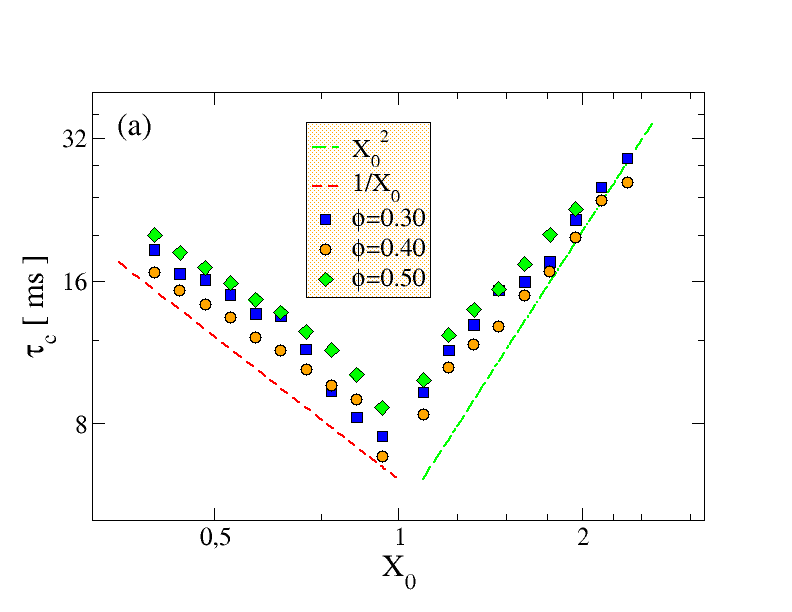}
\includegraphics[width=0.49\linewidth]{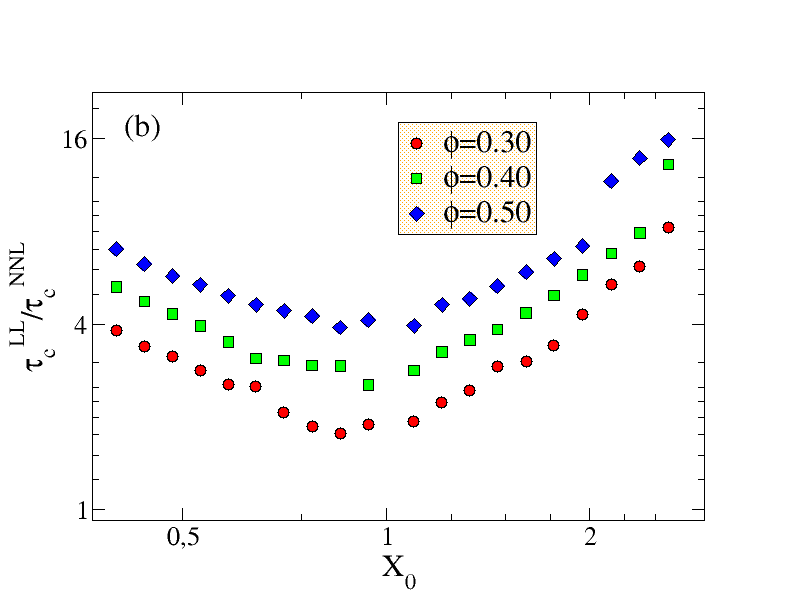}
\caption{\label{fig:CollTimeX0} 
Dependence of average collisione time $\tau_c$ on elongation $X_0$.
(a) $\tau_c$ at $\phi=0.30,0.40$ and $0.50$ using only LLs. 
In this case the average collision time is linked to the number of particles in the first neighbour shell 
and grows with elongation. The dashed lines are guides to the eyes showing the $1/X_0$
and $X_0^2$ behaviors (see text).
(b) Comparison of performance results with and without the use of NNLs plotting 
$\tau_c^{LL}/\tau_c^{NNL}$ versus the aspect ratio $X_0$ for $3$
different volume fractions  $\phi=0.30,0.40,0.50$.  $\tau_c^{NNL}$ is the collision time
employing NNLs and  $\tau_c^{LL}$ is the collision time using only LLs.  The use of LLs severely degrades 
the performance of the algorithm for elongated HEs.}
\end{center}
\end{figure}

In the following we will show results for simulations performed at volume fractions $\phi=0.30,0.40,0.50$.
First, the case where only LLs are employed is considered.
Besides the expected decrease of $\tau_c$ upon increasing the density, a marked 
increase of $\tau_c$ upon increasing/decreasing the elongation with respect to the hard-sphere case $X_0=1$ is clearly apparent in Fig. \ref{fig:CollTimeX0}(a). 
The explanation for such behavior is straightforward: as noticed in \cite{Donev05b}, the number of HEs in a LL cell increases as $X_0^2$ ($1/X_0$) for the prolate (oblate) case, and in turns
$\tau_c$ increases as $X_0^2$ ($1/X_0$) at big (small) elongations. 
In contrast, using NNLs $\tau_c$ is quite $X_0$-independent, increasing 
significantly the performance at big and small elongations with respect to LLs 
(see Fig. \ref{fig:CollTimeX0}(b)). 
In this case the number of neighbors is independent from $\phi$ on changing 
the elongation and the computational efficiency depends only on NR method efficiency,  that is quite $X_0$-indipendent.
On passing we note that at very high elongation ( $X_0 \gtrapprox 5$), the
NR method requires more steps to converge and the exact number of steps depends on the initial guess. Nevertheless, we checked that, at least up to $X_0=10$, the algorithm's
performance does not change (i.e.  $\tau_c$ remains $X_0$-independent) if NNLs are used.
 

\subsection{Optimizing the parameter $\epsilon_d$}\label{sec:epsdopt}

\begin{figure}
\includegraphics[width=0.425\linewidth]{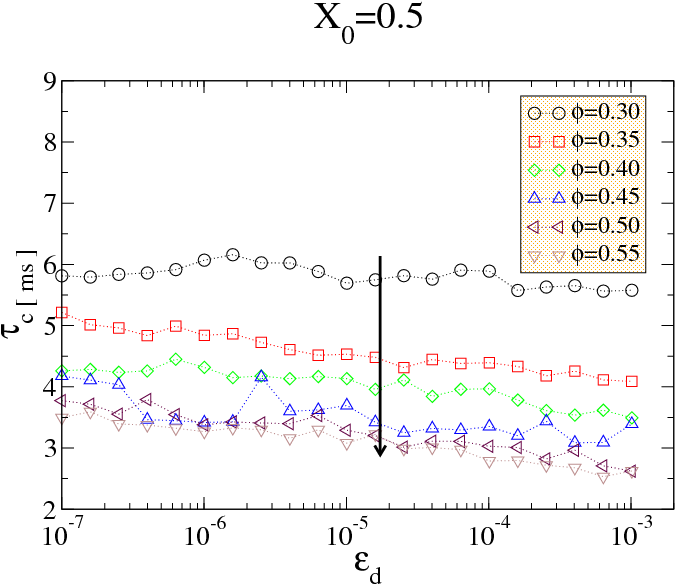}
\hskip 0.4cm
\includegraphics[width=0.425\linewidth]{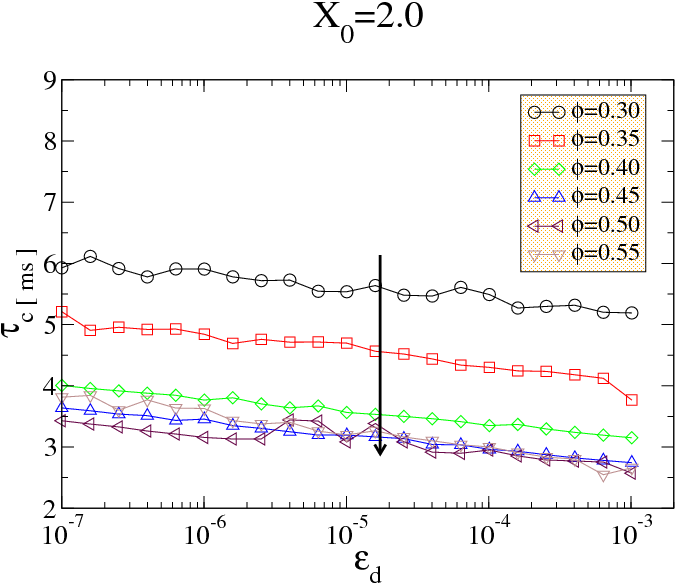}
\caption{\label{fig:epsd}
Dependence of $\tau_c$ on $\epsilon_d$ (using NNLs) for two different elongations ($X_0=0.5, 2.0$)
at various values of $\phi$ ranging from $0.30$ to $0.55$. 
Arrows indicate decreasing volume fractions.
}
\end{figure}

The main parameter entering in the  collision prediction algorithm   is $\epsilon_d$ (sec. \ref{sssec:bracket_time}) and it is important
to establish the optimal value in terms of efficiency. 
As already discussed in \ref{sssec:bracket_time}, too large a value for this parameter
may result in a failure of EDMD due to overlaps of HEs.  Hence, the best choice for $\epsilon_d$
is the largest value not generating HEs overlaps. 
 Fig. \ref{fig:epsd} shows that $\tau_c$ monotonically decreases
as $\epsilon_d$ is decreased, and that $\phi$-dependence is rather weak.
In practice, a value  like $\epsilon_d=10^{-4}-10^{-5}$ is usually a good, reliable and safe
choice for most situations.


\subsection{Optimizing neighbour lists}\label{sec:nnlopt}
\begin{figure}
\begin{center}
\includegraphics[width=0.425\linewidth]{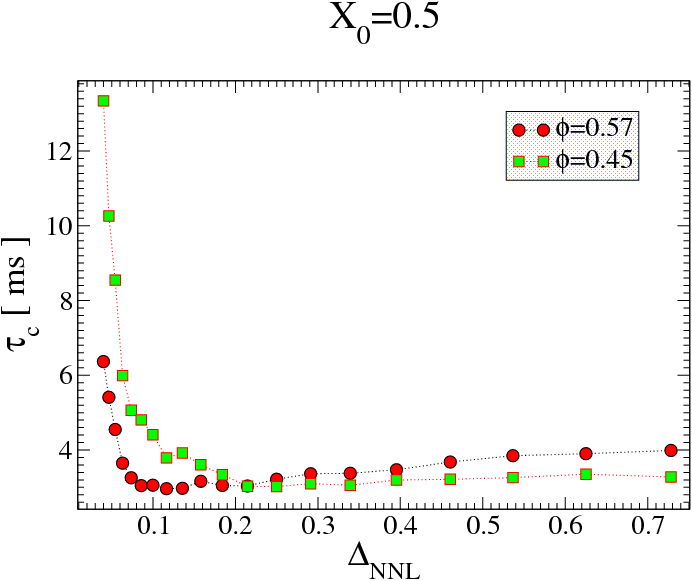}
\hskip 0.4cm
\includegraphics[width=0.425\linewidth]{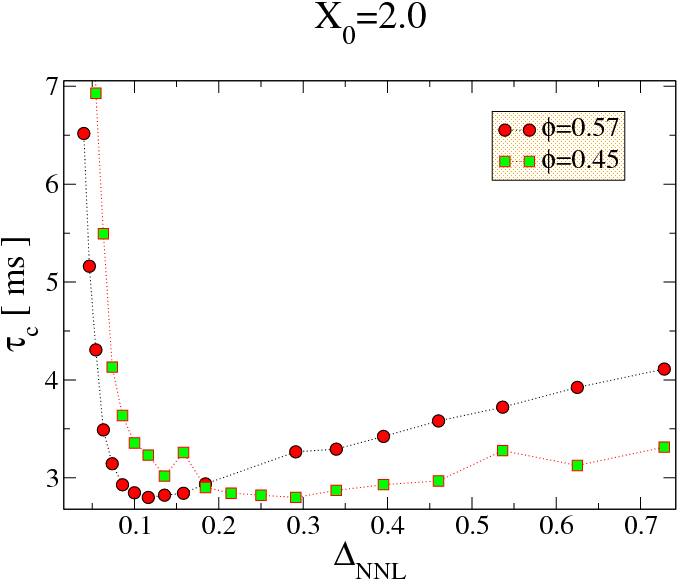}
\caption{\label{fig:delta}
The average collision time depends non-monotonically upon $\Delta_{NNL}$,
showing a clear minimum for a certain $\Delta_{NNL}^*$ value (see text). 
In this figure such a dependence is shown for $4$ particular state points ($X_0=0.5$ at $\phi=0.45, 0.57$
and $X_0=2.0$ at $\phi=0.45, 0.57$).
}
\end{center}
\end{figure}

\begin{figure}
\begin{center}
\includegraphics[width=0.425\linewidth]{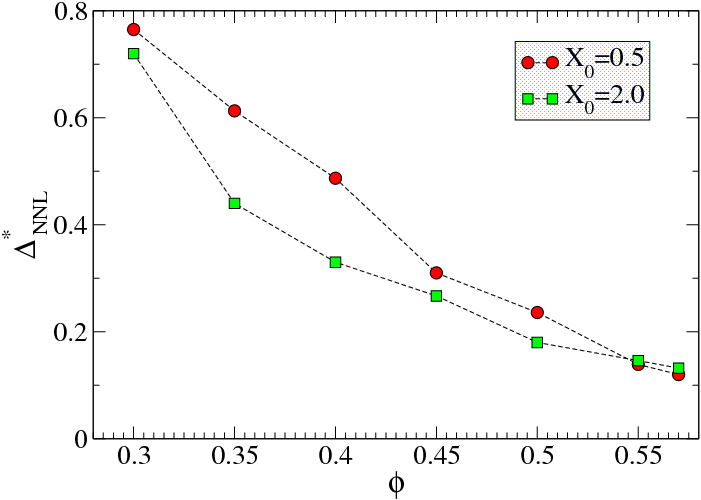}
\hskip 0.4cm
\includegraphics[width=0.425\linewidth]{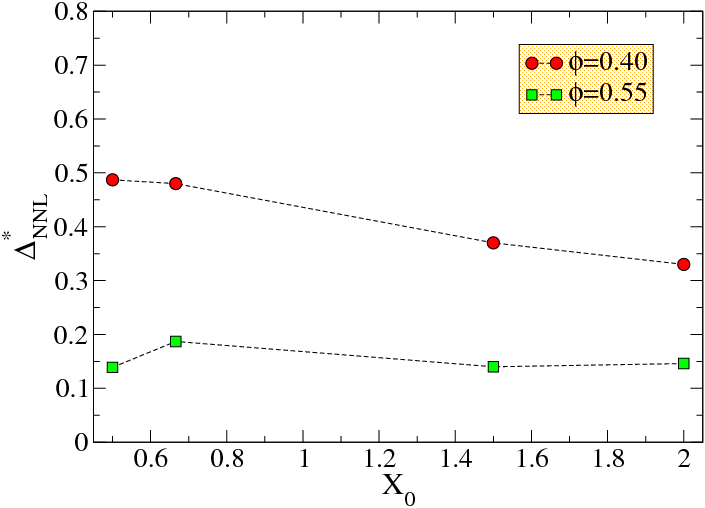}
\caption{\label{fig:optdelta}
In this figure the dependance of $\Delta^*_{NNL}$  on $\phi$ for  $X_0=0.5, 2.0$ 
(left panel) and the dependance of $\Delta^*_{NNL}$ on $X_0$ for $\phi=0.40, 0.55$ (right panel)
is shown.
$\Delta^*_{NNL}$ decreases significantly with increasing $\phi$ (left panel) but it has a weak dependence 
on $X_0$ (right panel). 
}
\end{center}
\end{figure}

Changing the parameter $\Delta_{NNL}$, i.e. changing the size of the bounding
boxes, computational times vary non-monotonically.
If $\Delta_{NNL}$ tends to $0$, the number of neighbors $N_{pc}^A$, $N_{pc}^B$ in Eq.(\ref{Eq:timepercollSimpl}) tends to $0$ accordingly, but NNLs updates tend to be infinitely 
frequent, because escape time of HE from their bounding boxes tends to $0$.
Therefore $\tau_c$ diverges to infinity, if $\Delta_{NNL}\rightarrow 0$.
If $\Delta_{NNL}\rightarrow \infty$ (assuming one has a system with infinite particles),  
escape times tends to $0$ and  time intervals between two successive NNLs rebuilds 
diverge, but $N_{pc}^A\rightarrow\infty$, $N_{pc}^B\rightarrow\infty$ in Eq.(\ref{Eq:timepercollSimpl}), so that again $\tau_c\rightarrow\infty$.
As a result, there must be a minimum of $\tau_c$ as a function of $\Delta_{NNL}$.
The value of $\Delta_{NNL}$, which minimizes $\tau_c$ is the optimal value
for this parameter and it will be labeled $\Delta^*_{NNL}$. Fig. \ref{fig:delta} shows for two volume fractions that $\tau_c (\Delta_{NNL})$ exhibits a clear minimum.
$\Delta^*_{NNL}$ depends much more on the volume fraction $\phi$ than on the elongation $X_0$ and this is clear from Fig. \ref{fig:optdelta}, where $\Delta^*_{NNL}$ is plotted as a function of $\phi$ for different $X_0$.
The latter result is due to the fact that in Eq. (\ref{Eq:timepercollSimpl})   
$\Delta t_{coll}$, through $N_{pc}^A$, $N_{pc}^B$, strongly depends on $\phi$, while
both $\Delta t_{coll}$ and $\Delta t_u$ weakly depend on $X_0$.


 
Fig. \ref{fig:optdelta} provides  a simple estimate of $\Delta^*_{NNL}$, that ranges from $0.1$ to $0.8$ 
if $0.5 < X_0 < 2.0$ and $0.25 < \phi < 0.55 $, upon changing volume fraction and elongation.

 

\section{Simulating Superquadrics.}\label{sec:supquad}
In the Sections \ref{sec:EMD4HE} and \ref{sec:perform} we have shown in detail how to apply the algorithm illustrated in Sec. \ref{sec:EMD4RB} to simulate HEs and its performance in this particular case.
The algorithm proposed by Donev {et al.} \cite{Donev05a,Donev05b} has been generalized to arbitrary convex shapes  \cite{DonevThesis06,DonevPRB06,JiaoPRL08,JiaoPRE09}, analogously we consider here the simulation of SQs, that are a possible generalization of HEs studied in the previous section, skipping anyway some details that will be supplied in a future publication.
    
\subsection{Geometry and Simulation Details}
\label{ssec:SQgeom}
The surface of an HE has been defined through Eq. (\ref{Eq:ellsurf}), that 
with a suitable choice of the reference system axes (principal axes) takes
the form:
\begin{equation}
\left (\frac{x}{a}\right )^2 + \left (\frac{y}{b}\right )^2 + \left (\frac{z}{c}\right )^2 = 1
\label{eq:HEexplicit}
\end{equation}
A straightforward generalization of the latter equation leads
to the definition of particles, called SQs, whose surface is the following:
\begin{equation}
f(x,y,z) = \left |\frac{x}{a}\right |^n + \left |\frac{y}{b}\right|^m + \left |\frac{z}{c}\right |^p - 1 = 0
\label{eq:SQdef}
\end{equation}
where the parameters $n,m,p$ are real numbers and $a$, $b$, $c$ are the SQ semi-axes. 
A monodisperse system of $N=512$ SQs has been simulated with $m=n=2$
and with two equal semi-axes, i.e. $b=c$.

\begin{figure}
\begin{center}
\includegraphics[width=0.85\linewidth]{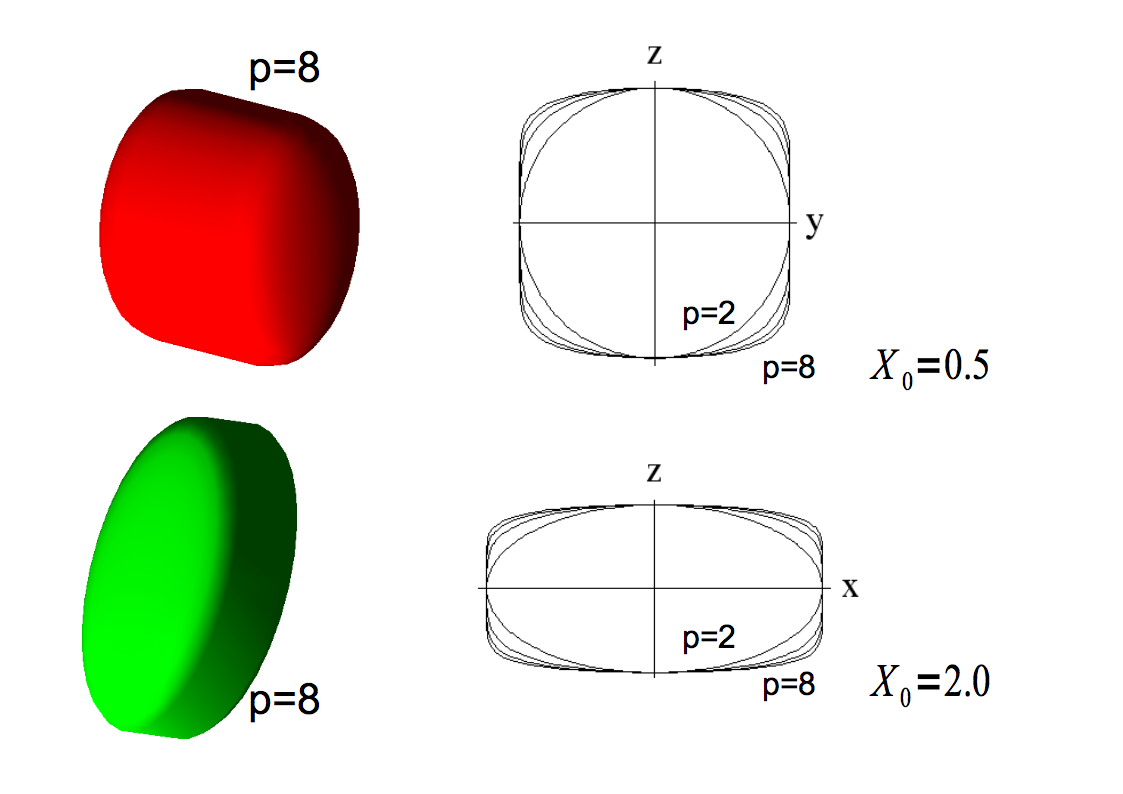}
\caption{\label{fig:SQshape}
Shape of the superquadrics used in the simulations. Top: on the left a prolate SQ
with $p=8$ and $X_0=0.5$ is represented in 3D. On the right the plots of the contour resulting from the intersection 
of the SQ (for several $p$ ranging from $2$ to $8$) with the plane $x=0$ are shown.
Bottom:  on the left an oblate SQ  with $p=8$ and $X_0=2.0$ is represented in 3D. 
On the right the plots of the contour resulting from the intersection of the SQ (for several $p$ ranging from $2$ to $8$) with the plane $y=0$ are shown.
}
\end{center}
\end{figure}

Such SQs can be characterized by the elongation $X_0=a/b$ and by the parameter
$p$, that determines the sharpness of the edges (see Fig. \ref{fig:SQshape}).
SQs of elongation $X_0 < 1$ are called ``oblate'', while SQs of elongation $X_0 > 1$
are called ``prolate'', as for the HEs.
Elongations $X_0=0.5$ and $X_0=2.0$ have been studied.
Nicely for $X_0=0.5$ and $p = 8$  the SQ has a pillow-like shape, while for $X_0=2.0$ and $p = 8$
it has a boat-like shape, as shown in Fig. \ref{fig:SQshape}.  
The system of prolate and oblate SQs has been simulated in a cubic box of volume $V$ with periodic boundary conditions 
at the volume fraction $\phi=0.256$ and at various values of $p$.
The length of the smallest semi-axes is chosen as $0.8$ times the unit of distance, the mass of the SQ as unit of mass ($m=1$) 
and the moment of inertia is spherically symmetric, i.e.  $I_x=I_y=I_z=1$, as in the case of HEs.

The code for simulating SQs can be straightforwardly derived from the HE code, indeed
for implementing the EDMD of SQs it suffices to:
\begin{itemize}
\item Evaluate the Jacobians in Eqs. (\ref{Eq:dist8}) ( or Eqs. (\ref{Eq:dist5}) ) and (\ref{Eq:jaccontime}) 
         using Eq. (\ref{eq:SQdef}). 
\item Calculate the Jacobians pertaining to the collision of a SQ with a plane
         to make use of NNL (see Sec. \ref{sec:NNL}).
\item Adapt the steepest descent method described in Sec. \ref{sec:iniguess}. 
\item Adapt the guess of the distance shown in Sec. \ref{ssec:HomoGuess}.
\end{itemize}

\begin{figure}
\begin{center}
\includegraphics[width=0.49\linewidth]{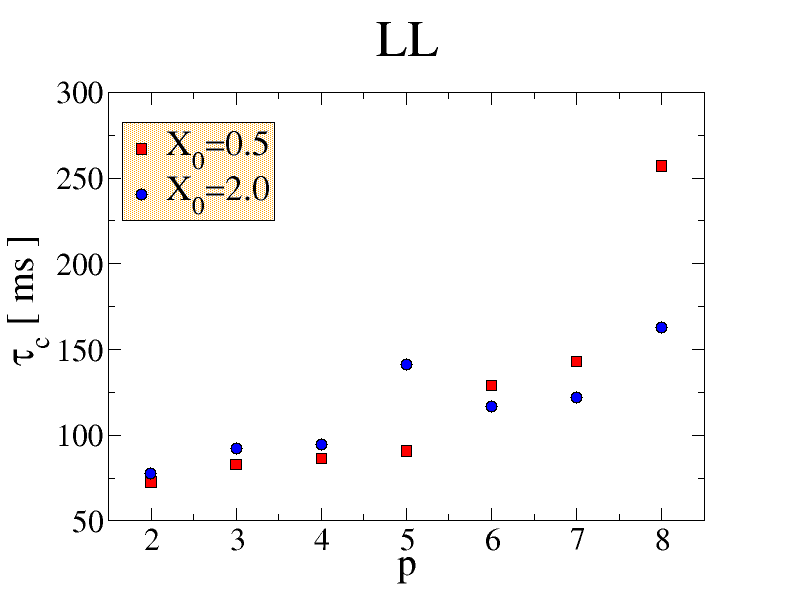}
\includegraphics[width=0.49\linewidth]{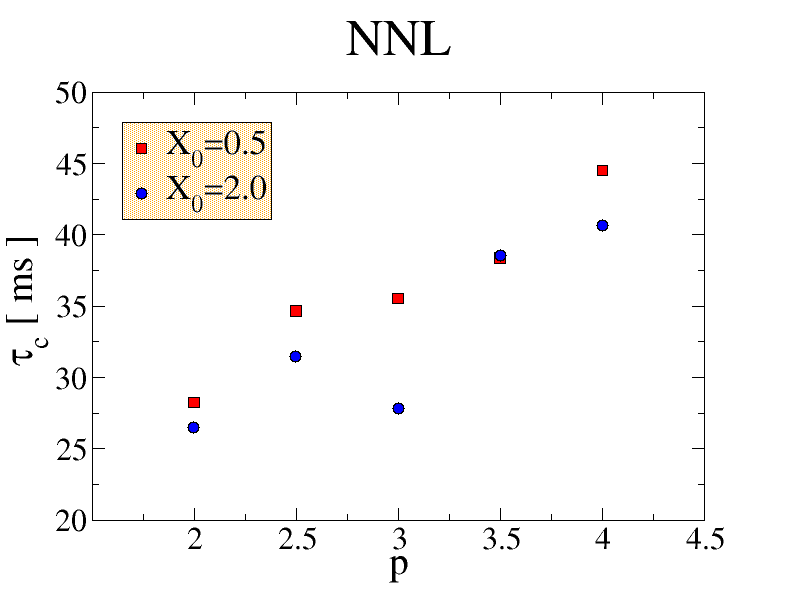}
\caption{\label{fig:SQsims}
Average collision time $\tau_c$ for simulations of $N=512$ SQs at $\phi=0.256$ 
for elongations $X_0=0.5$ and $X_0=2.0$ using only LL (left panel) 
and also using NNL (right panel).
}
\end{center}
\end{figure}
\subsection{Performance results}
\label{ssec:perfSQ}
Fig. \ref{fig:SQsims} shows the average collision time $\tau_c$ for the system of SQs using either  
LLs or NNLs.
With regard to the dependence of $\tau_c$ on $\phi$, $X_0$, $\epsilon_d$ and $\Delta_{NNL}$ 
We do not expect a significantly different behavior from what has been observed for HEs. 
Hence here we focus on the dependence of the average collision time on the parameter $p$.

For the LLs case $p$ ranges from $2$ to $8$, while enabling NNLs it ranges from $2$ to $4$ only.
In the case of LLs for values of $p > 8$ the NR for calculating the distance
between SQs starts having convergence problems, while in the case of NNLs
the NR for calculating the distance between a SQ and a plane starts having convergence issues. 
In the present implementation of the algorithm for locating the collision time between SQs and between 
a SQ and a plane the greater the $p$ the slower will be the NR method to reach convergence 
(up to a maximum value above which the NR method starts failing) and the greater will be the 
number of steps needed to locate the contact point (see Sec. \ref{sssec:bracket_time}). 


About the performance comparison between HEs and SQs from Fig. \ref{fig:SQsims} it turns out 
that the SQs simulations (see $p=2$, i.e. HEs) are at least $5$-$6$ times slower than the HEs ones
and this is mainly due to the more time consuming calculation of the Jacobians in Eqs (\ref{Eq:dist8}), (\ref{Eq:dist5}) and 
(\ref{Eq:jaccontime}).
Finally, as discussed above it is apparent from Fig.  \ref{fig:SQsims}  the increase of $\tau_c$ at high $p$ in all cases.

\section{Possible applications of the present algorithm}
\label{Sec:application}
The algorithm illustrated in this paper has been already applied in several fields, like biophysics, soft-matter and condensed-matter physics.   First of all, this new algorithm was employed to investigate the dynamics of HEs \cite{DeMicheleNEMGLASS}, finding, for the first time, evidence of a pre-nematic order-driven glass transition.
The fact that this algorithm can simulate hard objects of convex arbitrary shape offers new intriguing possibilities for investigating the changes in the phase diagram of HEs on changing particles shape.
For example, concerning the pre-nematic order driven glass transition observed in 
\cite{DeMicheleNEMGLASS}, it is stimulating to think about the possibility
to completely inhibit the nematic transition with a different choice of the HRB 
shape and/or introducing shape polydispersity. Work in this direction is under way.
Changing the HRB shape may result in a phase diagram, which is richer than the one of simple HEs, since
new phases (cubatic, smectic, ecc.) may appear \cite{FrenkelMulderMolPhys85}.
Moreover, changing the shape and introducing shape polydispersity may have
effects on packing and on mixing/demixing of hard objects.


This algorithm has also been extended \cite{DeMicheleSticky06} to combinations of hard and
localized attractive interactions, modeled as site-site square well potentials. In this way,
the hard-body may be decorated with an arbitrary number of spherical patches (sticky spots) arranged in fixed site locations. The algorithm for the prediction of the collision time between "sticky spots" is an adaptation of the algorithm illustrated in the present paper for HEs (some details can be found in \cite{DeMicheleSticky06} but this topic 
will be discussed extensively in a forthcoming publication).

Spheres decorated with sticky spots have been used to simulate primitive models
of water \cite{DeMicheleSticky06} and silica \cite{DeMicheleSilica06} and to study
the kinetics of self-assembly of an idealized fluid model \cite{DeMicheleDouglas08}.
Hard ellipsoids with sticky spots have been employed to study a model of a chemical gel 
\cite{DeMicheleDGEBA08}.

HEs decorated with attractive interacting sites may also  be used to model complex hard molecules, retaining some degree of flexibility. In this approach, the loss in the details of the system is counterbalanced by its extreme flexibility and by the possibility of investigating time scales which are not accessible by standard methods (like ``ab-initio'' calculations or full-atom molecular dynamics simulations). There are several directions in which this methodology could be applied. A very promising field of interest is represented by biophysical systems.
A particular problem that has received much attention over the last years is the IgG antibody-antigen interaction. An immunoglobulin (or antibody) is a ``Y'' shaped protein that is ubiquitous in most vertebrates. Its role is to bind to viruses or bacteria to facilitate their neutralization process. It is extremely interesting that a single object has the ability to bind to biological targets whose size can vary from much smaller to much larger than its own.
The introduction of square well interactions \cite{DeMicheleSticky06}  between hard-body particles can be used to generalize the Go model \cite{Go83} and  better account for steric effects between different proteins.
Previously, primitive model of proteins had been developed to study biophysical problems. In the $4$-beads model, described in \cite{buldyrev4beads}, each amino acid is represented by a maximum of four beads. Three beads correspond to the amide $N$, the $\alpha$-carbon $C$, the carbonyl $C'$ groups.
The fourth bead models the amino-acid side-chain group of atoms, and it is placed at the centre of the nominal $C_\beta$ atom.
Exploiting the new algorithm proposed in the present paper, each amino-acid  with its interacting (active) sites can be modeled as a set of spherical spots that form one unique rigid body. In biological simulations it is worth noting that the solvent can be also introduced explicitly in several ways:
\begin{itemize}
\item using a primitive model, e.g. the primitive model of water used in \cite{DeMicheleSticky06} to take into account the directionality of the hydrogen bonding.
\item using a simplified solvent of hard spherical particles (this technique has been used
for studying the interaction of IgG and antigenes)
\item adapting the method for performing brownian dynamics of square-well particles discussed in \cite{BD4HS} to the present algorithm.
\end{itemize}

Finally, this algorithm may find applications in the field of computer science: animations and virtual reality may benefit from the physical accuracy and efficiency of the present method.
In general, virtual motions of objects in 3D space require an accurate prediction of collisions and are fundamental for robotics, computer graphics, 3D computer games \cite{Eberly01} and CAD applications. For example, because of its flexibility in shape,  ellipsoids are often chosen as bounding volumes for robotic arms in collision detection \cite{Ju01,Rimon97,Wu96}.

\section{Conclusions}
\label{Sec:conclusions}

In this paper a new efficient method for performing event-driven molecular dynamics simulations of non-spherical HRBs has been proposed.
This new method is based on the traditional NR method for solving a set of non-linear equations.
In particular, geometrical distance and contact point and time between two moving HRBs can be calculated through the NR method in a very efficient way.
The method has been tested against a well established and studied system, the hard ellipsoids of revolution and also against a less common system like the SQs.  
Anyhow HEs and SQs are just particular cases for the present algorithm, since more general convex hard bodies can be simulated.
Furthermore also non-convex hard bodies can be in principle simulated by the present algorithm, 
if all the possible solutions of Eqs. (\ref{Eq:dist8}) (or Eqs. (\ref{Eq:dist5})) can be found, 
because in this case the actual distance can be simply obtained by Eq. (\ref{eq:distdef}).    

In the specific case of uniaxial HEs the time per collision achieved at moderate and high densities for elongations $X_0=0.5,2.0$ is around $2.5$ ms, a value which is comparable to the performance of the method described in \cite{Donev05b}.
For comparison, in our implementation the time per sphere collision is about $0.25$ ms, i.e. simulating hard spheres
is about an order of magnitude faster than simulating HEs, even for nearly spherical HEs (the same observation 
has been carried out in \cite{Donev05b}). 

To solve the set of equations to evaluate the contact point and time (see Eqs. (\ref{Eq:contime})) by the NR method,
a good initial guess (``bracketing") of the solution has to be provided.
It has been shown (see \ref{sssec:bracket_time}) that an initial guess for Eqs.  (\ref{Eq:contime}) can be found by evaluating the geometrical distance between the two colliding HRBs and by making use of a simple overestimate of the rate of variation of the distance ($\dot d_{max}$).  
It is worth stressing that using such an algorithm for bracketing the contact point and time,  ``grazing collisions'' 
do not constitute a problem within machine accuracy.
If $\epsilon_d \lessapprox 10^{-4}$ all collisions are correctly predicted and this choice of $\epsilon_d$ does not depend on volume fraction or aspect ratio of simulated HEs.
Also a new method to implement NNL based on oriented bounding boxes has been presented.
This new NNL method is fast and flexible enough to be easily extended to more complex shapes than simple HEs and SQs.

The present algorithm can also be generalized, without losing numerical efficiency, to the case of
attractive interactions modeled via discontinuous step-wise potentials, including the interesting case
of site-site square well interaction, providing the possibility of modeling highly directional interactions
between the rigid bodies.  This possibility can be further refined by transforming 
site-site interactions  into a permanent link between objects, by simply changing the depth
of square well interactions. In this way, the objects can be linked into flexible structures.
Such flexibility couples well with the novel NNL implementation described.

\section{Acknowledgments}

The author acknowledges support from CASPUR and Cofin.
The author warmly thanks Prof. F.~Sciortino, Dr. G. Foffi, F. Romano and A. De Michele for their careful reading of the manuscript, 
Prof. P.~Tartaglia for his ongoing support and Prof. G. Ciccotti for helpful discussions. The author also thanks Dr. A. Scala for useful discussions and for having taken part to the very early stages of this project.


\end{document}